\documentclass[]{jpconf}
\usepackage{graphicx}
\usepackage{amssymb}
\usepackage{amsmath, amscd}
\newcommand{\R}{\mathbb{R}}
\newcommand{\Z}{\mathbb{Z}}
\newcommand{\C}{\mathbb{C}}

\usepackage{lipsum}

\begin{document}

\title{Mathematics of the classical and the quantum}

\author{Alexey A. Kryukov}

\address{Department of Mathematics \& Natural Sciences, University of Wisconsin-Milwaukee, USA}

\ead{kryukov@uwm.edu}

\begin{abstract}
Newtonian and Schr{\"o}dinger dynamics can be formulated in a physically meaningful way within the same Hilbert space framework. This fact was recently used to discover an unexpected relation between classical and quantum motions that goes beyond the results provided by the Ehrenfest theorem. The Newtonian dynamics was shown to be the Schr{\"o}dinger dynamics of states constrained to a submanifold of the space of states, identified with the classical phase space of the system. Quantum observables are identified with vector fields on the space of states. The commutators of observables are expressed through the curvature of the space. The resulting embedding of the Newtonian and Schr{\"o}dinger dynamics into a unified geometric framework is rigid in the sense that the Schr{\"o}dinger dynamics is a unique extension of the Newtonian one. Under the embedding, the normal distribution of measurement results associated with a classical measurement implies the Born rule for the probability of transition of quantum states. The mathematics of the discovered relationship between the classical and the quantum is reviewed and investigated here in detail, and applied to the process of measurement of spin and position observables. 

\end{abstract}


\section{Introduction}

In a recent series of papers \cite{KryukovMath}-\cite{KryukovMacro1}, an important new connection between the classical and quantum dynamics was derived. The starting point was a realization of classical and quantum mechanics on an equal footing within the same Hilbert space framework and identification of observables with vector fields on the sphere of normalized states. 
This resulted in a physically meaningful interpretation of components of the velocity of state. Newtonian dynamics was shown to be the Schr{\"o}dinger dynamics of a system whose state is constrained to the classical phase space submanifold in the Hilbert space of states. 
In simple words, the classical space and classical phase space of a system of particles can be identified with a submanifold of the space of states of the corresponding quantum system. When the system is constrained to the submanifold, it behaves classically. Otherwise, it behaves quantum-mechanically. The velocity of the state at any point of the classical space submanifold can be decomposed into classical (velocity, acceleration) and non-classical (phase velocity, spreading) components.  The curvature of the sphere of states is determined from the canonical commutation relations.

These results suggest that there is an alternative approach to quantum mechanics that is more appropriate for understanding and visualizing the theory and for addressing its fundamental problems and paradoxes. In this paper, the mathematics of such an approach will be presented and applied to investigating the process of measurement in classical and quantum physics. After the background information, a relationship between the normal probability distribution, typical for classical measurements, and the Born rule for transition of quantum states will be derived. This relationship is then illustrated and its dynamical origin is revealed.
It will be argued that by accepting the space of states as a new arena for physical events and identifying the classical space and classical phase space with submanifolds of thereof we can fruitfully explore the relationship of the classical and quantum dynamics, including the process of measurement, in a coherent and fundamentally simple way.

\section{Newtonian mechanics in the Hilbert space of states} 
\label{Newton}

Everyday experience shows that macroscopic bodies possess a well-defined position in space at any moment of time. In quantum mechanics, the state of a spinless particle with a known position ${\bf a}$ is given by the Dirac delta function $\delta^{3}_{\bf a}({\bf x})=\delta^{3}({\bf x}-{\bf a})$. The map $\omega: {\bf a} \longrightarrow \delta^{3}_{\bf a}$ provides a one-to-one correspondence between points ${\bf a} \in \R^{3}$ and state ``functions" $\delta^{3}_{\bf a}$. This allows us to describe points in $\R^3$ in functional terms and identify  the set $\R^{3}$ with the set $M_{3}$ of all delta functions in the space of state functions of the particle. 

Dirac delta states are considered an idealization. But so is the notion of a material point in Newtonian mechanics. Both idealizations are the building blocks in their respective theories. As we will see, they are also important for understanding the relationship between Newtonian physics and quantum mechanics. We will see that Newtonian physics in the Euclidean space $\R^3$ is the Schr{\"o}dinger quantum mechanics of systems whose state is constrained to the submanifold in the Hilbert space of states, formed by the delta-like states of particles.

The space $L_{2}({\R}^{3})$ does not contain delta functions. For instance, if $f_{n}$ is a delta-convergent sequence \cite{Gelfand} of continuous, square-integrable functions on $\R^3$, then the sequence $\int f^2_{n}({\bf x})d^3{\bf x}$ diverges. There are essentially two ways out of this difficulty. One method is to approximate delta functions with the more physical Gaussian functions, which are in  $L_{2}({\R}^{3})$. Another one is to complete the Hilbert space  $L_{2}({\R}^{3})$ to obtain a wider space that includes delta functions. The methods are essentially equivalent and will be used interchangeably.
To explain, let us write the inner product of functions $\varphi, \psi \in L_{2}(\R^{3})$ as
\begin{equation}
\label{innerdd}
(\varphi, \psi)_{L_{2}}=\int \delta^{3}({\bf x}-{\bf y})\varphi({\bf x}){\overline \psi}({\bf y})d^{3}{\bf x}d^{3}{\bf y},
\end{equation}
where $\delta^{3}({\bf x}-{\bf y})$ is the kernel of the identity operator.
By approximating $\delta^{3}({\bf x}-{\bf y})$  with a Gaussian function, one obtains a new inner product in  $L_{2}(\R^3)$
\begin{equation}
\label{hh}
(\varphi, \psi)_{\bf H}=\int e^{-\frac{({\bf x}-{\bf y})^{2}}{8\sigma^{2}}}\varphi({\bf x})\overline{\psi}({\bf y})d^{3}{\bf x}d^{3}{\bf y}.
\end{equation}
Here $\sigma$ is a parameter.
The Hilbert space ${\bf H}$ obtained by completing $L_{2}(\R^3)$ with respect to this inner product contains delta functions and their derivatives.
In particular,
\begin{equation}
\label{delta1}
\int e^{-\frac{({\bf x}-{\bf y})^{2}}{8\sigma^{2}}}\delta^{3}({\bf x}-{\bf a})\delta^{3}({\bf y}-{\bf a})d^{3}{\bf x}d^{3}{\bf y}=1.
\end{equation}
Furthermore, the injective map $\omega$ is continuous and is, in fact, a homeomorphism onto the image $\omega(\R^{3})$ with the topology induced by the metric on ${\bf H}$: two delta functions $\delta^3_{\bf a}, \delta^3_{\bf b}$ are close in ${\bf H}$ if and only if ${\bf a }$ and ${\bf b}$ are close in $\R^3$. Furthermore, $\omega$ and its inverse are smooth.
It follows that the set $M_{3}$ of all delta functions  $\delta^{3}_{\bf a}({\bf x})$ with ${\bf a} \in \R^{3}$ form a submanifold of the unit sphere in the Hilbert space ${\bf H}$, diffeomorphic to $\R^3$. The map $\omega: {\bf a} \longrightarrow \delta^{3}_{\bf a}$ becomes an embedding of $\R^3$ into ${\bf H}$.

The map $\rho_{\sigma}: {\bf H} \longrightarrow L_{2}(\R^{3})$ that relates $L_{2}$ and ${\bf H}$-representations and identifies the two methods of dealing with delta-states is given by the Gaussian kernel
\begin{equation}
\label{sigma}
\rho_{\sigma}({\bf x},{\bf y})=\left(\frac{1}{2\pi \sigma^{2}}\right)^{3/4}e^{-\frac{({\bf x}-{\bf y})^{2}}{4\sigma^{2}}}.
\end{equation}
In fact, it is easy to see that $\rho_{\sigma}$ is one-to-one. Indeed, taking various derivatives of $(\rho_{\sigma}f)({\bf x})$ one can see that all Fourier coefficients of $f$ in the basis of (multivariable) Hermite functions in $L_{2}(\R^3)$ vanish. Since these functions form a basis in $L_{2}(\R^3)$, we conclude that $f=0$, hence, $\rho_{\sigma}$  is one-to-one.
Multiplying the operators (integrating the product of kernels) one can see that
\begin{equation}
\label{GG}
k({\bf x}, {\bf y})=(\rho^{\ast}_{\sigma}\rho_{\sigma})({\bf x}, {\bf y})=e^{-\frac{({\bf x}-{\bf y})^{2}}{8\sigma^{2}}},
\end{equation}
which is consistent with (\ref{hh}) and proves that $\rho_{\sigma}$ is an isomorphism of the Hilbert spaces $L_{2}(\R^3)$ and ${\bf H}$. 

The isomorphism $\rho_{\sigma}$ transforms delta functions $\delta^{3}_{\bf a}$ to Gaussian functions ${\widetilde \delta^{3}_{\bf a}}=\rho_{\sigma}(\delta^{3}_{\bf a})$, centered at ${\bf a}$. The image $M^{\sigma}_{3}$ of $M_{3}$ under $\rho_{\sigma}$ is an embedded submanifold of the unit sphere in $L_{2}(\R^{3})$ made of the functions ${\widetilde \delta^{3}_{\bf a}}$. The map $\omega_{\sigma}=\rho_{\sigma} \circ \omega: \R^3 \longrightarrow M^{\sigma}_{3}$ is a diffeomorphism. Here $\omega$ is the same as before.  Note that the kernel $\delta^{3}({\bf x}-{\bf y})$ of the metric on $L_{2}({\R}^{3})$ is analogous to the Kronecker delta $\delta_{ik}$, representing the Euclidean metric in orthogonal coordinates. The ``skewed" kernel $e^{-\frac{({\bf x}-{\bf y})^{2}}{8\sigma^{2}}}$ of the metric on ${\bf H}$ is then analogous to a constant non-diagonal matrix $g_{ik}$ representing the Euclidean metric in linear coordinates with skewed axes.

Let ${\bf r}={\bf a}(t)$ be a path with values in $\R^{3}$ and let $\varphi_{t}=\delta^{3}_{{\bf a}(t)}$ be the corresponding path in $M_{3}$. Integration by parts in (\ref{hh}) on this path results in the following expression for the speed of motion in ${\bf H}$:
\begin{equation}
\label{parts1}
\left \| \frac{d \varphi}{dt} \right \|^{2}_{H}=
\left.\frac {\partial^{2}k({\bf x},{\bf y})}{\partial x^{i} \partial y^{k}}\right|_{{\bf x}={\bf y}={\bf a}} \frac {d{\bf a}^{i}}{dt}\frac {d{\bf a}^{k}}{dt}.
\end{equation}
Here  $k({\bf x},{\bf y})=e^{-\frac{({\bf x}-{\bf y})^{2}}{8\sigma^{2}}}$ as in (\ref{GG}), so that
\begin{equation}
\left.\frac {\partial^{2}k({\bf x},{\bf y})}{\partial x^{i} \partial y^{k}}\right|_{{\bf x}={\bf y}={\bf a}}=\frac{1}{4\sigma^{2}}\delta_{ik},
\end{equation}
where $\delta_{ik}$ is the Kronecker delta symbol. Assuming now that the distance in $\R^{3}$ is measured in the units of $2\sigma$, we obtain 
\begin{equation}
\label{Norms}
\left \| \frac{d \varphi}{dt} \right \|_{H}=\left \| \frac{d {\bf a}}{dt} \right \|_{\R^{3}}.
\end{equation}
It follows that the map $\omega: \R^{3} \longrightarrow {\bf H}$ is an isometric embedding. Furthermore, the set $M_{3}$ is complete in ${\bf H}$ so that there is no vector in ${\bf H}$ orthogonal to all of $M_{3}$. In fact, if $(f, \delta^{3}_{\bf a})_{\bf H}=0$, then $\rho_{\sigma}(f)=0$ and so $f=0$, because $\rho_{\sigma}$ is an isomorphism.

By defining the operations of addition $\oplus$ and multiplication by a scalar $\lambda \odot$  via $\omega({\bf a})\oplus\omega({\bf b})=\omega({\bf a}+{\bf b})$ and $\lambda \odot\omega({\bf a})=\omega(\lambda {\bf a})$ with $\omega$ as before, we obtain $M_{3}$ as a vector space isomorphic to the Euclidean space $\R^{3}$. Since $\omega$ is an embedding, these operations are continuous in the topology of ${\bf H}$. Of course, the obtained vector structure on $M_{3}$  is not the same as the one on the Hilbert space ${\bf H}$ and $M_{3}$ is not a subspace of ${\bf H}$.

With the classical space in place, we can now proceed with a reformulation of Newtonian mechanics in functional terms. The projection of velocity and acceleration of the state $\delta^{3}_{{\bf a}(t)}$ onto the Euclidean space $M_{3}$ yields correct Newtonian velocity and acceleration of the classical particle:
\begin{equation}
\label{v1}
\left( \frac{d}{dt}\delta^{3}_{{\bf a}}({\bf x}), -\frac{\partial}{\partial x^{i}} \delta^{3}_{{\bf a}}({\bf x})\right)_{\bf H} =\frac{d a^{i}}{dt}
\end{equation}
and 
\begin{equation}
\label{a11}
\left( \frac{d^{2}}{dt^{2}}\delta^{3}_{{\bf a}}({\bf x}) , -\frac{\partial}{\partial x^{i}} \delta^{3}_{{\bf a}}({\bf x})\right)_{\bf H} =\frac{d^{2} a^{i}}{dt^{2}}.
\end{equation}
These equations follow from the chain rule
\begin{equation}
\frac{d}{dt}\delta^{3}({\bf x}-{\bf a})= -\frac{\partial}{\partial x^{i}} \delta^{3}({\bf x}-{\bf a})\frac{d a^{i}}{dt}
\end{equation}
and the integration by parts in the inner products in (\ref{v1}) and (\ref{a11}). 

The Newtonian dynamics of the classical particle can be derived from the principle of least action for the action functional $S$ on paths in ${\bf H}$, defined by
\begin{equation}
\label{action}
S=\int k({\bf x},{\bf y})\left[\frac{m}{2}\frac{d \varphi_{t}({\bf x})}{dt} \frac{d{\overline  \varphi_{t}}({\bf y})}{dt}-V({\bf x}) \varphi_{t}({\bf x}) {\overline \varphi_{t}}({\bf y})\right]d^{3}{\bf x}d^{3}{\bf y}dt.
\end{equation}
Here $m$ is the mass of the particle, $V$ is the potential and $k({\bf x}, {\bf y})=e^{-\frac{1}{2}({\bf x}-{\bf y})^{2}}$, as in (\ref{GG}) with $2\sigma=1$, to ensure (\ref{Norms}). In fact, under the constraint $\varphi_{t}({\bf x})=\delta^{3}({\bf x}-{\bf a}(t))$ the action (\ref{action}) becomes
\begin{equation}
\label{aa}
S=\int\left[\frac{m}{2}\left(\frac{d{\bf a}}{dt}\right)^{2}-V({\bf a})\right]dt,
\end{equation}
which is the classical action functional for the particle. An action functional for the time-dependent Schr{\"o}dinger equation that reduces to the classical action (\ref{aa}) on the properly constrained  states will be introduced in section \ref{components}.

It follows that a classical particle can be considered a constrained dynamical system with the state $\varphi$ of the particle and the velocity of the state $\frac{d\varphi}{dt}$  as dynamical variables. 
A similar realization exists for classical mechanical systems consisting of any number of particles. For example, the map $\omega \otimes \omega: \R^{3}\times \R^{3} \longrightarrow {\bf H}\otimes {\bf H}$, $\omega \otimes \omega ({\bf a}, {\bf b})=\delta^{3}_{\bf a} \otimes \delta^{3}_{\bf b}$ identifies the configuration space $\R^{3}\times \R^{3}$ of a two particle system with the embedded submanifold $M_{6}=\omega \otimes \omega(\R^{3}\times \R^{3})$ of the Hilbert space ${\bf H}\otimes {\bf H}$. Consider a path $({\bf a}(t), {\bf b}(t))$ in $\R^{3}\times \R^{3}$ and the corresponding path $\delta^{3}_{{\bf a}(t)}\otimes \delta^{3}_{{\bf b}(t)}$ with values in $M_{6}$. For any $t$, the vectors $\frac{d}{dt}\delta^{3}_{{\bf a}(t)}\otimes \delta^{3}_{{\bf b}(t)}$ and $\delta^{3}_{{\bf a}(t)}\otimes \frac{d}{dt}\delta^{3}_{{\bf b}(t)}$ are tangent to $M_{6}$ at the point $\delta^{3}_{{\bf a}(t)}\otimes \delta^{3}_{{\bf b}(t)}$ and orthogonal in ${\bf H}\otimes {\bf H}$. The space $M_{6}$ with the induced metric is isometric to the direct product $\R^{3}\times \R^{3}$ with the natural Euclidean metric. Projection of velocity and acceleration of the state $\varphi(t)=\delta^{3}_{{\bf a}(t)}\otimes \delta^{3}_{{\bf b}(t)}$ onto the basis vectors $\left(- \frac{\partial}{\partial x^i}\delta^{3}_{{\bf a}(t)}\right)\otimes \delta^{3}_{{\bf b}(t)}$ and $\delta^{3}_{{\bf a}(t)}\otimes \left(-\frac{\partial}{\partial x^{k}}\delta^{3}_{{\bf b}(t)}\right)$ yields the velocity and acceleration of the particles by means of the formulas similar to (\ref{v1}) and (\ref{a11}).

\section{Observables as vector fields}

Quantum observables can be identified with vector fields on the space of states. Given a self-adjoint operator ${\widehat A}$ on a Hilbert space $L_{2}$ of square-integrable functions (it could in particular be the tensor product space of a many body problem)  one can introduce the associated linear vector field $A_{\varphi}$ on $L_{2}$ by
\begin{equation}
\label{vector}
A_{\varphi}=-i{\widehat A}\varphi.
\end{equation}
If $D$ is the domain of the operator ${\widehat A}$, then $A_{\varphi}$ maps $D$ into the vector space $L_{2}$.
Because ${\widehat A}$ is self-adjoint, the field $A_{\varphi}$, being restricted to the sphere $S^{L_{2}}$ of unit normalized states, is tangent to the sphere.
The commutator of observables and the commutator (Lie bracket) of the corresponding vector fields are related in a simple way:
\begin{equation}
\label{comm}
[A_{\varphi},B_{\varphi}]=[{\widehat A},{\widehat B}]\varphi.
\end{equation}

Furthermore, a Hilbert metric on the space of states yields a Riemannian metric on the sphere. For this, consider the realization $L_{2\R}$ of the Hilbert space $L_{2}$, i.e., the real vector space of pairs $X=(\mathrm{Re} \psi, \mathrm{Im} \psi)$ with $\psi$ in $L_{2}$. If $\xi, \eta$ are vector fields on $S^{L_{2}}$, define a Riemannian metric $G_{\varphi}: T_{\R\varphi}S^{L_{2}}\times T_{\R\varphi}S^{L_{2}} \longrightarrow \R$ on the sphere by
\begin{equation}
\label{Riem}
G_{\varphi}(X,Y)=\mathrm{Re} (\xi, \eta).
\end{equation}
Here $X=(\mathrm{Re} \xi, \mathrm{Im} \xi)$, $Y=(\mathrm{Re} \eta, \mathrm{Im} \eta)$ and  $(\xi, \eta)$ denotes the $L_{2}$-inner product of $\xi, \eta$. 

The Riemannian metric on $S^{L_{2}}$ yields a Riemannian (Fubini-Study) metric on the projective space $CP^{L_{2}}$, which is the base of the fibration $\pi:S^{L_{2}} \longrightarrow CP^{L_{2}}$. For this, an arbitrary tangent vector $X \in T_{R\varphi}S^{L_{2}}$ is decomposed into two components: tangent and orthogonal to the fibre $\{\varphi\}$ through $\varphi$ (i.e., to the plane $C^{1}$ containing the circle $S^{1}=\{\varphi\}$). The differential $d\pi$ maps the tangential component to the zero-vector. The orthogonal component of $X$ can be then identified with $d\pi(X)$. 
If two vectors $X,Y$ are orthogonal to the fibre $\{\varphi\}$, the inner product of $d\pi(X)$ and $d\pi(Y)$ in the Fubini-Study metric is equal to the inner product of $X$ and $Y$ in the metric $G_{\varphi}$:
\begin{equation}
(d\pi(X), d\pi(Y))_{FS}= G_{\varphi}(X,Y).
\end{equation}

The resulting metrics can be used to find physically meaningful components of vector fields  $A_{\varphi}$ associated with observables. 
Since $A_{\varphi}$ is tangent to $S^{L_{2}}$, it can be decomposed into components tangent and orthogonal to the fibre $\{\varphi\}$ of the fibre bundle $\pi: S^{L_{2}} \longrightarrow CP^{L_{2}}$. These components have a simple physical meaning, justifying the use of the projective space $CP^{L_{2}}$.
From
\begin{equation}
{\overline A} \equiv (\varphi, {\widehat A}\varphi)=(-i\varphi, -i{\widehat A}\varphi),
\end{equation}
one can see that the expected value of an observable ${\widehat A}$ in state $\varphi$ is the projection of the vector $-i{\widehat A}\varphi \in T_{\varphi}S^{L_{2}}$ onto the fibre $\{\varphi\}$.
Because
\begin{equation}
(\varphi, {\widehat A}^{2}\varphi)=({\widehat A}\varphi, {\widehat A}\varphi)=(-i{\widehat A}\varphi, -i{\widehat A}\varphi),
\end{equation}
the term $(\varphi, {\widehat A}^{2}\varphi)$ is the norm of the vector $-i{\widehat A}\varphi$ squared. The vector $-i{\widehat A}_{\bot}\varphi=-i{\widehat A}\varphi-(-i{\overline A}\varphi)$ associated with the operator ${\widehat A}-{\overline A}I$ is  orthogonal to the fibre $\{\varphi\}$.
Accordingly, the variance 
\begin{equation}
\label{uncert1}
\Delta A^{2}=(\varphi, ({\widehat A}-{\overline A}I)^{2}\varphi)=(\varphi, {\widehat A}_{\bot}^{2}\varphi)=(-i{\widehat A}_{\bot}\varphi, -i{\widehat A}_{\bot}\varphi) 
\end{equation}
is the norm squared of the component $-i{\widehat A}_{\bot}\varphi$. 
Recall that the image of this vector under $d\pi$ can be identified with the vector itself. 
It follows that the norm of $-i{\widehat A}_{\bot}\varphi$ in the Fubini-Study metric coincides with its norm in the Riemannian metric on $S^{L_{2}}$ and in the original $L_{2}$-metric. 

The Schr{\"o}dinger equation
\begin{equation}
\label{evoll}
\frac{d\varphi}{dt}=-i{\widehat h}\varphi
\end{equation}
is an equation for the integral curves of the vector field $-i{\widehat h}\varphi$ on the sphere $S^{L_{2}}$.
Let's decompose $-i{\widehat h}\varphi$ onto the components parallel  and orthogonal to the fibre. The parallel component of $\frac{d\varphi}{dt}$ is numerically
\begin{equation}
{\mathrm Re} (-i\varphi, -i{\widehat h} \varphi)={\overline E},
\end{equation}
i.e., the expected value of the energy. The decomposition of the velocity vector $\frac{d\varphi}{dt}$ into the parallel and orthogonal components is then given by 
\begin{equation}
\label{evolll1}
\frac{d\varphi}{dt}=-i{\overline E}\varphi+-i({\widehat h}-{\overline E})\varphi=
-i{\overline E}\varphi-i {\widehat h}_{\perp}\varphi.
\end{equation}
By considering the orthogonal component of (\ref{evolll1}), we see that the orthogonal component of the velocity  $\frac{d\varphi}{dt}$ is equal to $-i {\widehat h}_{\perp}\varphi$. From this and equation (\ref{uncert1}) we conclude that: 
{\it The speed of evolution of state in the projective space is equal to the uncertainty of energy.}
Equation (\ref{evolll1}) also demonstrates that the physical state is driven by the operator  ${\widehat h}_{\perp}$, associated with the uncertainty in energy rather than the energy itself.

The realization of operators by vector fields yields other interesting results. For instance, the uncertainty relation
\begin{equation}
\label{uncert}
\Delta A \Delta B \ge \frac{1}{2}\left |\left(\varphi, [{\widehat A},{\widehat B}]\varphi\right)\right|
\end{equation}
follows geometrically from the comparison of areas of rectangle $A_{|XY|}$ and parallelogram  $A_{XY}$ formed by vectors $X=-i{\widehat A}_{\bot}\varphi$ and $Y=-i{\widehat B}_{\bot}\varphi$:
\begin{equation}
\label{obv}
A_{|XY|} \ge A_{XY}.
\end{equation}
There is also an uncertainty identity, \cite{KryukovUncert}:
\begin{equation}
\label{Pyth}
\Delta A^{2} \Delta B^{2}=A^{2}_{XY}+ G^{2}_{\varphi} (X, Y).
\end{equation}
The sum on the right hand side of (\ref{Pyth}) can be written as $||X||^2||Y||^2 \sin^2 \theta +  ||X||^2||Y||^2 \cos^2 \theta$, where $\theta$ is the angle between $X$ and $Y$. 
In particular, when $\theta=0$, the uncertainty comes from the inner product term $G_{\varphi} (X, Y)$ in (\ref{Pyth}) and when $\theta = \pi/2$, the uncertainty is due to the area term. By replacing ${\widehat B}$ with a real linear combination of the operators ${\widehat A}, {\widehat B}$ (i.e., by rotating $B_{\varphi}$ in the plane through $X$ and $Y$), we can change $\theta$ in any desirable way while preserving the product $\Delta A^{2} \Delta B^{2}$.

\section{Commutator of observables and curvature of the sphere of states}
\label{curvature}

The identification of observables with vector fields allows one to relate the commutators of observables with the curvature of the sphere of states.
To see this, consider first the space 
 $\C^{2}$ of electron's spin states. 
The sphere $S^{3}$ of unit-normalized states in $\C^{2}$ can be identified with the group manifold $SU(2)$.  
For this, one identifies the space $\C^{2}$ of complex vectors $\varphi=\left[ \begin{array}{c}
z_{1} \\ 
z_{2}
\end{array}
\right]$ 
with the space $M$ of $2 \times 2$ matrices 
\begin{equation}
\label{MatN}
{\widehat \varphi}=\left[ 
\begin{array}{cc}
z_{1} & z_{2} \\ 
-{\overline z}_{2} & {\overline z}_{1}
\end{array}
\right].
\end{equation}
The map $\widehat{\omega}: \varphi \longrightarrow {\widehat \varphi}$ is an isomorphism of (real) vector spaces $\C^{2}$ and $M$.
The sphere $S^{3}$ of unit states in $\C^{2}$ is identified via $\widehat{\omega}$ with the subset of matrices with unit determinant. The latter subset is the group $SU(2)$ under matrix multiplication.

The differential $d\widehat{\omega}$ of the map $\widehat{\omega}$ identifies the tangent space $T_{e_{1}}S^{3}$ to the sphere $S^{3}$ at the point
$e_{1}=\left[ 
\begin{array}{c}
1  \\ 
0
\end{array}
\right]$ (that is, the hyperplane $\mathrm{Re} z_{1}=1$)
with the Lie algebra $su(2)$ of traceless anti-Hermitian matrices 
\begin{equation}
\label{Lie}
{\widehat A}=\left[ 
\begin{array}{cc}
ia_{2} & a_{3}+ia_{4} \\ 
-a_{3}+ia_{4} & -ia_{2}
\end{array}
\right],
\end{equation}
 $a_{2}, a_3, a_{4} \in R$.
Under $d\widehat{\omega}$ the basis 
$e_{2}=\left[ 
\begin{array}{c}
i  \\ 
0
\end{array}
\right]$,
$e_{3}=\left[ 
\begin{array}{c}
0  \\ 
1
\end{array}
\right]$,
$e_{4}=\left[ 
\begin{array}{c}
0  \\ 
i
\end{array}
\right]$ in the tangent space $T_{e_{1}}S^{3}=R^{3}$ becomes the basis $\{i{\widehat \sigma}_{3}, i{\widehat \sigma}_{2}, i{\widehat \sigma}_{1}\}$ in the Lie algebra $su(2)$. In particular, the real numbers $a_{2}, a_{3}, a_{4}$ acquire the meaning of coordinates of points on the tangent space $\mathrm{Re} z_{1}=1$ in the basis $\{e_{2}, e_{3}, e_{4}\}$.

The embedding of $S^{3}$ into $\C^{2}$ induces the usual Riemannian metric on the sphere. A direct verification demonstrates that this metric coincides with the Killing metric on $SU(2)$. The latter metric can be defined on the tangent space $T_{ e}SU(2)$ at the identity $e$ (i.e., on the Lie algebra $su(2)$) by $({\widehat X}, {\widehat Y})_{K}=\frac{1}{2}Tr {\widehat X} {\widehat Y}^{+}$ and then extended to the entire $SU(2)$ by the group action. Here $({\widehat X}, {\widehat Y})_{K}$ denotes the Killing inner product of tangent vectors and ${\widehat Y}^{+}$ on the right is the Hermitian conjugate of ${\widehat Y}$. The constant $1/2$ in the Killing metric together with a proper choice of the unit of measurement ensure the equality of the Riemannian and the Killing metrics. The tangent space $su(2)$ is spanned by the spin operators having the dimension of angular momentum and measured in the units of $\hbar$. Therefore, the Planck system of units will be used. The spin generators ${\widehat s}_{1}=\frac{i}{2} {\widehat \sigma}_{1}, {\widehat s}_{2}=\frac{i}{2}{\widehat \sigma}_{2}, {\widehat s}_{3}=\frac{i}{2}{\widehat \sigma}_{3}$ are orthogonal in the defined metric and have a norm equal to $1/2$ in Planck units. 

The integral curves of the left-invariant vector fields 
$L_{\widehat X}({\widehat \varphi})={\widehat \varphi}{\widehat X}$ are geodesics on $SU(2)$. They are given by 
${\widehat \varphi}_{t}={\widehat \varphi}_{0}e^{-i{\widehat X}t}$. In the usual coordinates on $\C^2$,  the equation of these geodesics takes the form $\varphi_{t}=e^{-i{\widehat X}t}\varphi_{0}$, where $\omega(\varphi_0)={\widehat \varphi}_{0}$. The carriers of geodesics are the great circles on the sphere $S^{3}$.
The commutators of the spin observables are directly related to the sectional curvature of the sphere $S^{3}$. This is not surprising as the non-trivial Lie bracket of vector fields whose integral curves are geodesics can only be due to curvature of the underlying space. If ${\widehat X}, {\widehat Y} \in su(2)$ are linearly independent generators and $L_{\widehat X}({\widehat \varphi}), L_{\widehat Y}({\widehat \varphi})$ are the associated left-invariant vector fields, then the sectional curvature $R_{\varphi}(p)$ of $S^{3}$ in the plane $p$ through $L_{\widehat X}({\widehat \varphi}), L_{\widehat Y}({\widehat \varphi})$ is given at any point ${\widehat \varphi}$ by 
\begin{equation}
\label{normmN}
R_{\varphi}(p)=\frac{1}{4}\frac{\left\|[{\widehat X}, {\widehat Y}]\right\|^{2}_{K}}{\left\|{\widehat X}\right\|^{2}_{K}\left\|{\widehat Y}\right\|^{2}_{K}-\left({\widehat X}, {\widehat Y}\right)^2_{K}}.
\end{equation}
In particular, if the generators ${\widehat X}, {\widehat Y}$ are orthonormal in the Killing metric, (\ref{normmN}) simplifies to
\begin{equation}
\label{normmN1}
R_{\varphi}(p)=\frac{1}{4}\left\|[{\widehat X}, {\widehat Y}]\right\|^{2}_{K}.
\end{equation}

Using the formula (\ref{normmN}), we obtain the following expression for the sectional curvature 
$R_{\varphi}(p)$ in the plane $p$ through orthogonal vectors $L_{{\widehat s}_{1}}({\widehat \varphi}),L_{{\widehat s}_{2}}({\widehat \varphi})$: 
\begin{equation}
\label{section}
R_{\varphi}(p)=\frac{1}{4} \frac{ \left([{\widehat s_{1}},{\widehat s_{2}}],[{\widehat s_{1}},{\widehat s_{2}}]\right)_{K}}{\left({\widehat s}_{1},{\widehat s}_{1}\right)_{K}\left({\widehat s}_{2},{\widehat s}_{2}\right)_{K}}=4\left ({\widehat s}_{3},{\widehat s}_{3}\right)_{K}=1.
\end{equation}
This means that the radius of $S^{3}$ in Planck units is equal to $1$, confirming the isometric nature of the isomorphism $\widehat{\omega}$ considered as a map from the unit sphere $S^3$ in $\C^2$ onto $SU(2)$ with the Killing metric. 
Note that in an arbitrary system of units the sectional curvature would be equal to $1/ \hbar^{2}$ (i.e., radius=$\hbar$). The dimension of sectional curvature is consistent with the fact that the tangent space $su(2)$ is spanned by the spin operators.

The relation obtained between commutators of spin observables and radius of the sphere of states can be extended to other observables. In particular, the commutator $[{\widehat p},{\widehat x}]$ of position and momentum observables of an arbitrary non-relativistic particle with states in the space $L_{2}(\R)$  yields similarly the sectional curvature of the sphere $S^{L_{2}}$ in $L_{2}(\R)$.
In fact, let's compute the sectional curvature of the sphere $S^{L_{2}}$ in the plane through the tangent vectors $-i{\widehat p}\varphi$, $-i{\widehat x}\varphi$  at a point $\varphi \in S^{L_{2}}$.
It is convenient to represent the action of operators ${\widehat p}$, ${\widehat x}$ in the basis $\varphi_{n}(x)=\frac{1}{\sqrt [4]{\pi}2^{n}n!}H_{n}(x)e^{-\frac{x^{2}}{2}}$, $n=0, 1, 2, ...$ of the quantum harmonic oscillator. Here $H_{n}(x)$ are the Hermite polynomials. Note that the vectors $\varphi_{n}$ are in the domain of the operators ${\widehat p}$, ${\widehat x}$, ${\widehat p}{\widehat x}$ and ${\widehat x}{\widehat p}$. The matrices of the operators ${\widehat p}$, ${\widehat x}$ in the basis are given by
\begin{equation}
\label{mat1}
{\widehat x}=\frac{1}{\sqrt {2}}\left[ 
\begin{array}{ccccc}
0 & 1 & 0 & 0 & \cdot \cdot \cdot\\ 
1 & 0 & {\sqrt 2} & 0 & \cdot \cdot \cdot\\
0 & {\sqrt 2} & 0 & {\sqrt 3} & \cdot \cdot \cdot\\
0 & 0 & {\sqrt 3} & 0 & \cdot \cdot \cdot\\
\cdot & \cdot & \cdot & \cdot & \cdot \cdot \cdot
\end{array}
\right]
\end{equation}
and
\begin{equation}
\label{mat2}
{\widehat p}=\frac{1}{\sqrt {2}}\left[ 
\begin{array}{ccccc}
0 & -i & 0 & 0 & \cdot \cdot \cdot\\ 
i & 0 & -i{\sqrt 2} & 0 & \cdot \cdot \cdot\\
0 & i{\sqrt 2} & 0 & -i{\sqrt 3} & \cdot \cdot \cdot\\
0 & 0 & i{\sqrt 3} & 0 & \cdot \cdot \cdot\\
\cdot & \cdot & \cdot & \cdot & \cdot \cdot \cdot
\end{array}
\right].
\end{equation}
Because the operators ${\widehat x}$, ${\widehat p}$ are unbounded, the validity of such a matrix representation requires a discussion. However, for the purpose of computing the sectional curvature it will be sufficient to point out that the matrices (\ref{mat1}) and (\ref{mat2}) correctly reproduce the action of operators on all vectors with finitely many non-vanishing components in the basis $\{\varphi_{n}\}$.

Let us find the sectional curvature of the sphere $S^{L_{2}}$ at the ``vacuum'' point $\left.\varphi_{n}\right|_{n=0}=\varphi_{0}$. For this, consider the subspace $\C^{2} \subset L_{2}(\R)$ formed by the first two vectors of the basis.  
Note that up to the coefficient $\frac{1}{\sqrt {2}}$, the sub-matrices formed by the first two rows and columns of matrices (\ref{mat1}) and (\ref{mat2}) coincide with the Pauli matrices ${\widehat \sigma}_{x}, {\widehat \sigma}_{y}$ respectively. Let us introduce the bounded operators ${\widehat s}_{p}, {\widehat s}_{x}$ on $L_{2}(\R)$ defined by ${\widehat s}_{x}\varphi=\frac{1}{\sqrt{2}}{\widehat \sigma}_{x}\varphi$, ${\widehat s}_{p}\varphi=\frac{1}{\sqrt{2}}{\widehat \sigma}_{y}\varphi$ for $\varphi$ in $\C^{2}$, and by ${\widehat s}_{x}\varphi=0$, ${\widehat s}_{p}\varphi=0$ for $\varphi$ in the orthogonal complement of $\C^2$ in $L_{2}(\R)$.
Note that the action of operators ${\widehat p}$, ${\widehat x}$ and $[{\widehat p}$, ${\widehat x}]$ on the point $\varphi_{0}$ is correctly reproduced by the operators ${\widehat s}_{p}, {\widehat s}_{x}$:
\begin{eqnarray}
\label{xs}
{\widehat x}\varphi_{0}&=&{\widehat s}_{x}\varphi_{0} \\
\label{pps}
{\widehat p}\varphi_{0}&=&{\widehat s}_{p}\varphi_{0} \\
\label{compx}
 \left[{\widehat p}, {\widehat x}\right]\varphi_{0}&=&\left[{\widehat s}_{p}, {\widehat s}_{x}\right ]\varphi_{0}.
\end{eqnarray}

Consider the sphere $S^{3}=S^{L_{2}}\cap \C^{2}$ with the metric induced by the inclusion. 
As discussed, this metric coincides with the Killing metric on the group $SU(2)=S^{3}$. The point $\varphi_{0}$ is given in the basis $\{\varphi_{0}, \varphi_{1}\}$ in $\C^{2}$ by the column
$\left[ 
\begin{array}{c}
1 \\ 
0
\end{array}
\right]$.
The image ${\widehat \varphi}_{0}$ of the column $\varphi_{0}$ under the isomorphism (\ref{MatN}) 
is the identity $e$ in the group $SU(2)$. Accordingly, one can compute the norms of the right sides of (\ref{xs}), (\ref{pps}) and (\ref{compx}) in the Killing metric. Such a computation verifies that these norms are equal to the norms of the corresponding left sides in the $L_{2}$-metric. For example, the norm of the right side of (\ref{compx}) in the Killing metric is given by $\left\|{\widehat \varphi}_{0}\frac{1}{2}[{\widehat \sigma}_{y}, {\widehat \sigma}_{x}]\right\|_{K}= \left\|i{\widehat \sigma}_{z}\right\|_{K}=\sqrt {\frac{1}{2}Tr ({\widehat \sigma}_{z})^{2}}=1$. This coincides with the $L_{2}$-norm of the corresponding left side: $\left\|[{\widehat p}, {\widehat x}]\varphi_{0}\right\|_{L_{2}}=\left\|\varphi_{0}\right\|_{L_{2}}=1$.

The sectional curvature of $S^{L_{2}}$ in the plane through vector fields $-i{\widehat x}\varphi$, $-i{\widehat p}\varphi$ at $\varphi=\varphi_{0}$ is equal to the sectional curvature $R_{\varphi_{0}}(p)$ of $S^{3}$ in the plane $p$ through the fields $-i{\widehat \sigma}_{x}\varphi$, $-i{\widehat \sigma}_{y}\varphi$ at this point.
By (\ref{normmN}), (\ref{xs}), (\ref{pps}) and (\ref{compx}), this sectional curvature
is given in terms of the Lie brakets of these fields, i.e., in terms of the commutator  $[{\widehat p}, {\widehat x}]$ evaluated at $\varphi_{0}$ and is equal to $1$. Because sphere has a constant sectional curvature, the same result applies to any point. 
It follows that the commutator of vector fields associated with the operators of position and momentum has the same geometric interpretation as the commutator of vector fields associated with the operators of spin. 
Namely, the commutators give the sectional curvature of the sphere of states and produce the same value $1$ ($\hbar$ in an arbitrary system of units) for the radius of the sphere. This provides one with a purely geometric approach to quantum observables and their commutators in terms of vector fields on the sphere of states and their Lie bracket. 

\section{Components of the velocity of state under the Schr{\"o}dinger evolution}
\label{components}

We now have all necessary ingredients to put the classical and quantum mechanics on an equal footing and to discover their innermost relationship.
From (\ref{evolll1}), we know that for any state $\varphi \in S^{L_{2}}$, the velocity of state $\frac{d\varphi}{dt}$ in the Schr{\"o}dinger equation can be decomposed onto the components parallel and orthogonal to the fibre $\{\varphi\}$ of the bundle $\pi: S^{L_{2}} \longrightarrow CP^{L_{2}}$:
\begin{equation}
\label{evolll}
\frac{d\varphi}{dt}=
-i{\overline E}\varphi-i {\widehat h}_{\perp}\varphi.
\end{equation}
The norm of the parallel component $-i{\overline E}\varphi$ is the expected value of energy ${\overline E}$. It represents the phase velocity of state. The norm of the orthogonal component $-i {\widehat h}_{\perp}\varphi$ is equal to the uncertainty of energy $\Delta E$ on the state $\varphi$.  It represents the velocity of motion of the fibre $\{\varphi\}$. In particular, from (\ref{evolll}) it follows that under the Schr{\"o}dinger evolution, the speed of evolution of state in the projective space is equal to the uncertainty in energy.

The orthogonal component $-i{\widehat h}_{\bot}\varphi$ of the velocity can be further decomposed into physically meaningful components.
To see this, let's begin with an equation that follows from the Schr{\"o}dinger dynamics:
\begin{equation}
\label{projj}
 \left(\frac{d  \varphi}{dt}, -i {\widehat A} \varphi \right)=
 \left( \varphi, \frac{1}{2}\{{\widehat A}, {\widehat h} \}\varphi \right)-\left(\varphi,\frac{1}{2}[{\widehat A}, {\widehat h}]\varphi\right).
\end{equation}
The left hand side of (\ref{projj}) is the projection of the velocity of state onto the vector field associated with the observable ${\widehat A}$.
The imaginary part of the projection (the term with the commutator $[{\widehat A}, {\widehat h}]$) yields the Ehrenfest theorem for a time-independent observable ${\widehat A}$. The real part of this projection (the term with the anticommutator $\{{\widehat A}, {\widehat h}\}$) is the projection in the sense of Riemannian metric on $S^{L_{2}}$. This Riemannian projection can be used to identify further components of the velocity of state.

Suppose that at $t=0$, a microscopic particle is prepared in the state
\begin{equation}
\label{initial}
    \varphi_{{\bf a},{\bf p}}({\bf x})=\left(\frac{1}{2\pi\sigma^{2}}\right)^{3/4}e^{-\tfrac{({\bf x}-{\bf a})^{2}}{4\sigma^{2}}}e^{i\tfrac{{\bf p}({\bf x}-{\bf a})}{\hbar}},
\end{equation}
where $\sigma$ is the same as in (\ref{sigma}) and ${\bf p}=m{\bf v}_{0}$ with ${\bf v}_{0}$ being the initial group-velocity of the packet. 
Consider the subset $M^{\sigma}_{3,3}$ of all initial states $\varphi_{{\bf a},{\bf p}}$ given by (\ref{initial}) in $L_{2}(\R^3)$.
The map $\Omega: \R^{3}\times \R^{3} \longrightarrow M^{\sigma}_{3,3}$,
\begin{equation}
\label{omega}
\Omega({\bf a},{\bf p})=   \varphi_{{\bf a},{\bf p}}({\bf x}),
\end{equation} 
is a homeomorphism from the classical phase space onto $M^{\sigma}_{3,3}$ with the topology induced by the metric on $L_{2}(\R^3)$. In fact, it is one-to-one and the points $({\bf a}, {\bf p})$ and $({\bf b}, {\bf q})$ are close in $\R^3 \times \R^3$ if and only if the functions 
$ \varphi_{{\bf a},{\bf p}}$, $ \varphi_{{\bf b},{\bf q}}$ are close in $L_{2}(\R^3)$. The map $\Omega$ and its inverse are also smooth, so that $M^{\sigma}_{3,3}$ is a $6$-dimensional embedded submanifold of $L_{2}(\R^3)$ diffeomorphic to the classical phase space. 

Consider the set of all fibres of the bundle $\pi: S^{L_{2}} \longrightarrow CP^{L_{2}}$ through the points of $M^{\sigma}_{3,3}$. The resulting bundle $\pi: M^{\sigma}_{3,3}\times S^{1} \longrightarrow M^{\sigma}_{3,3}$ identifies $M^{\sigma}_{3,3}$ with a submanifold of $CP^{L_{2}}$, denoted by the same symbol. 
For $\Omega({\bf a},{\bf p})=r e^{i\theta}$, where $r$ is the modulus and $\theta$ is the argument of $\Omega({\bf a},{\bf p})$, the vectors $\frac{\partial r}{\partial a^{\alpha}}e^{i\theta}$ and $i\frac{\partial \theta}{\partial p^{\beta}}re^{i\theta}$ are orthogonal in the Riemannian metric on the sphere $S^{L_{2}}$. They are also orthogonal to the fibre $\{\varphi_{{\bf a},{\bf p}}\}$ in $L_{2}(\R^3)$ and can be, therefore, identified with vectors tangent to the projective manifold $M^{\sigma}_{3,3}$ at $\{\varphi_{{\bf a},{\bf p}}\}$. The Riemannian metric induced on  $M^{\sigma}_{3,3}$ is the Fubini-Study metric on $CP^{L_{2}}$, constrained to $M^{\sigma}_{3,3}$.

For any path $\{\varphi\}=\{\varphi_\tau\}$ with values in $M^{\sigma}_{3,3}\subset CP^{L_{2}}$, the norm of velocity vector $\frac{d \{\varphi\}}{d\tau}$ in the Fubini-Study metric is given by
\begin{equation}
\label{phaseMetric}
\left\|\frac{d \{\varphi\}}{d \tau}\right\|^{2}_{FS}=\frac{1}{4\sigma^{2}}\left\|\frac{d {\bf a}}{d\tau}\right\|^{2}_{\R^{3}}+\frac{\sigma^{2}}{\hbar^{2}}\left\|\frac{d {\bf p}}{d\tau}\right\|^{2}_{\R^{3}}.
\end{equation}
It follows that under a proper choice of units, the map $\Omega$ is an isometry that identifies the Euclidean phase space $\R^{3}\times \R^{3}$ of the particle with the submanifold $M^{\sigma}_{3,3} \subset CP^{L_{2}}$ furnished with the induced Fubini-Study metric. The map $\Omega$ is an extension of the isometric embedding $\omega_{\sigma}=\rho_{\sigma}\circ\omega$ introduced in section \ref{Newton} from the classical space to the classical phase space.

The obtained embedding of the classical phase space into the space of quantum states is physically meaningful.
To see this, let us first project the orthogonal component $-\frac{i}{\hbar}{\widehat h}_{\perp}\varphi$ of the velocity $\frac{d\varphi}{dt}$ onto vectors tangent to the curves of constant values of ${\bf p}$ and ${\bf a}$ (classical space and momentum space components) in the projective manifold $M^{\sigma}_{3,3}$. 
Calculation of the projection of the velocity $\frac{d \varphi}{dt}$ onto the unit vector $-\widehat{\frac{\partial r}{\partial a^{\alpha}}}e^{i\theta}$ (i.e., the classical space component of $\frac{d\varphi}{dt}$) for an arbitrary Hamiltonian of the form ${\widehat h}=-\frac{\hbar^{2}}{2m}\Delta+V({\bf x})$ yields
\begin{equation}
\label{pproj}
\left.\mathrm{Re}\left(\frac{d \varphi}{dt}, -\widehat{ \frac{\partial r}{\partial a^{\alpha}}}e^{i\theta}\right)\right|_{t=0}=\left.\left(\frac{d r}{dt}, -\widehat{ \frac{\partial r}{\partial a^{\alpha}}}\right)\right|_{t=0}=\frac{v^{\alpha}_{0}}{2\sigma}.
\end{equation}
Calculation of the projection of velocity $\frac{d \varphi}{dt}$ onto the unit vector $i\widehat{\frac{\partial\theta}{\partial p^{\alpha}}}\varphi$ (momentum space component) gives
\begin{equation}
\label{w}
\left.\mathrm{Re} \left(\frac{d\varphi}{dt}, i\widehat{\frac{\partial\theta}{\partial p^{\alpha}}}\varphi\right)\right|_{t=0}=\frac{mw^{\alpha} \sigma}{\hbar},
\end{equation}
where
\begin{equation}
\label{A}
mw^{\alpha}=-\left.\frac{\partial V({\bf x})}{\partial x^{\alpha}}\right|_{{\bf x}={\bf x}_{0}}
\end{equation}
and $\sigma$ is assumed to be small enough for the linear approximation of $V({\bf x})$ to be valid within intervals of length $\sigma$. 

The velocity $\frac{d\varphi}{dt}$ also contains component due to the change in $\sigma$ (spreading), which is orthogonal to the fibre $\{\varphi\}$ and the phase space $M^{\sigma}_{3,3}$, and is equal to
\begin{equation}
\label{spreadcomp}
\mathrm{Re} \left (\frac{d\varphi}{dt}, i\widehat{\frac{d\varphi}{d\sigma}}\right)=\frac{\sqrt{2}\hbar}{8\sigma^{2}m}.
\end{equation}
Calculation of the norm of $\frac{d\varphi}{dt}=\frac{i}{\hbar}{\widehat h}\varphi$ at $t=0$ gives
\begin{equation}
\label{decomposition}
\left\|\frac{d\varphi}{dt}\right\|^{2}=\frac{{\overline E}^{2}}{\hbar^{2}}+\frac{{\bf v}^{2}_{0}}{4\sigma^{2}}+\frac{m^{2}{\bf w}^{2}{\sigma}^{2}}{\hbar^{2}}+\frac{\hbar^{2}}{32\sigma^{4}m^{2}},
\end{equation}
which is the sum of squares of the found components. This completes a decomposition of the velocity of state at any point $\varphi_{{\bf a},{\bf p}} \in M^{\sigma}_{3,3}$. 

For a closed system, the norm  of $\frac{d\varphi}{dt}=\frac{i}{\hbar}{\widehat h}\varphi$ is preserved in time. For a system in a stationary state, this amounts to conservation of energy. In fact, in this case $\varphi_{t}({\bf x})=\psi({\bf x})e^{-\frac{iEt}{\hbar}}$, which is a motion along the phase circle, and
\begin{equation}
\left\| \frac{d\varphi}{dt}\right\|^{2}=\frac{E^2}{\hbar^2}.
\end{equation}
For a closed system in any initial state, the norm of the phase component (expected energy) and orthogonal component (energy uncertainty) of the velocity $\frac{d\varphi}{dt}$ are both preserved. 

In the linear potential approximation, valid in the considered case of small $\sigma$ (the choice of $\sigma$ is in our hands; the largest value of $\sigma$ consistent with observations is related to the boundary between classical and quantum), the first term in (\ref{decomposition}) is the square of the term
\begin{equation}
\label{restmass}
\frac{1}{\hbar}\left(U+K+\frac{\hbar^2}{8m\sigma^{2}}\right),
\end{equation}
where $U=V({\bf x}_{g})$ and $K=\frac{m{\bf v}^{2}_{g}}{2}$ are potential and kinetic energy of the packet considered as a particle with position ${\bf x}_{g}={\bf x}_{0}+{\bf v}_{0}t+\frac{{\bf w}t^{2}}{2}$ and velocity ${\bf v}_{g}={\bf v}_{0}+{\bf w}t$. The last term in parentheses in (\ref{restmass}) accounts for the difference in energy of the packets with the same $U$ and $K$, but different values of $\sigma$. Up to a constant factor, this term equals the component of velocity due to spreading given by (\ref{spreadcomp}). With the unit of length $2\sigma$ given by Compton length, this term is equal to half of the rest energy $mc^{2}$ of the particle, making it possible to identify the mass with the speed of motion of state due to spreading.

From (\ref{pproj}) and (\ref{w}), and a simple consistency check showing that the rate of change of the projection in (\ref{pproj}) is given by acceleration ${\bf w}$, one can see that the phase space components of the velocity of state $\frac{d\varphi}{dt}=-\frac{i}{\hbar}{\widehat h}\varphi$ assume correct classical values at any point $\varphi_{{\bf a},{\bf p}}  \in M^{\sigma}_{3,3}$. This remains true for the time dependent potentials as well. The immediate consequence of this and the linear nature of the Schr{\"o}dinger equation is that:
\begin{quote}
 \begin{em}
 Under the Schr{\"o}dinger evolution with the Hamiltonian ${\widehat h}=-\frac{\hbar^{2}}{2m}\Delta+V({\bf x},t)$, the state constrained to $M^{\sigma}_{3,3}\subset CP^{L_{2}}$ moves like a point in the phase space representing a particle in Newtonian dynamics. 
More generally, 
Newtonian dynamics of $n$ particles is the Schr{\"o}dinger dynamics of $n$-particle quantum system whose state is constrained to the phase-space submanifold $M^{\sigma}_{3n,3n}$ of the projective space for $L_{2}(\R^{3})\otimes \ ... \ \otimes L_{2}(\R^{3})$, formed by tensor product states $\varphi_{1}\otimes \ ... \ \otimes \varphi_{n}$ with $\varphi_{k}$ of the form (\ref{initial}). 
\end{em}
\end{quote}
Note again that the velocity and acceleration terms in (\ref{decomposition}) are orthogonal to the fibre $\{\varphi_{{\bf a},{\bf p}}\}$ of the fibration $\pi: S^{L_{2}}\longrightarrow CP^{L_{2}}$, showing that these Newtonian variables have to do with the motion in the projective space $CP^{L_{2}}$. The velocity of spreading is orthogonal to the fibre and to the phase space submanifold $M^{\sigma}_{3,3}$. Besides the derivation provided here, this can be seen directly from the symmetry properties of the terms of the time derivative of the state function that describes the spreading of a Gaussian wave packet.  The implication of this is that the ``concentration" of state under the collapse has nothing to do with a motion in the classical space.

Note that the functional 
\begin{equation}
S[\varphi]=\int \overline{\varphi}({\bf x}) \left[i\hbar \frac{\partial}{\partial t}-{\widehat h}\right] \varphi({\bf x}) d^3 {\bf x} dt
\end{equation}
 with ${\widehat h}=-\frac{\hbar^{2}}{2m}\Delta+V({\bf x},t)$ is the action functional for the Schr{\"o}dinger equation. At the same time, for the states $\varphi$ constrained to the manifold $M^{\sigma}_{3,3}$ this functional is equal to the classical action. Namely, for $\varphi$ varying over the states $\varphi_{{\bf a},{\bf p}}$ of the form (\ref{initial}), the action $S[\varphi]$ is equal to
\begin{equation}
S=\int \left[{\bf p}\frac{d {\bf a}}{dt}-h({\bf p},{\bf a})\right]dt,
\end{equation}
where $h({\bf p},{\bf a})=\frac{{\bf p}^2}{2m}+V({\bf a})+C$ is the Hamiltonian function and the constant $C$ is the ``rest energy" term in (\ref{restmass}). It follows that there exists a single action functional for the classical and quantum dynamics, which was perviously observed in a related context by John Klauder in \cite{enhanced}.

\section{Uniqueness of extension of Newtonian dynamics to  $CP^{L_{2}}$}
\label{unique}

The velocity of state under the Schr{\"o}dinger evolution with the Hamiltonian ${\widehat h}=-\frac{\hbar^{2}}{2m}\Delta+V({\bf x})$ was shown to contain for the states in $M^{\sigma}_{3,3}$ the classical velocity and acceleration  (formulae (\ref{pproj}) and (\ref{w})). This was used to establish that Newtonian dynamics of a particle is the Schr{\"o}dinger dynamics of the system whose state is constrained to the classical phase space $M^{\sigma}_{3,3}$. 

On the contrary, there exists a unique extension of the Newtonian dynamics formulated on the classical phase space $M^{\sigma}_{3,3}$ to a unitary dynamics in the Hilbert space $L_{2}(\R^3)$. 
More precisely: 
\begin{quote}
\begin{em}
Suppose that for any initial state  $\varphi_{{\bf a},{\bf p}}$ of the form
\begin{equation}
\label{del1}
\varphi_{{\bf a},{\bf p}}({\bf x})=
\left(\frac{1}{2\pi \sigma^{2}}\right)^{3/4}e^{-\frac{({\bf x}-{\bf a})^{2}}{4\sigma^{2}}}e^{i\frac{{\bf p}({\bf x}-{\bf a})}{\hbar}}
\end{equation}
there exists a path $\varphi=\varphi_{t}$  in $L_{2}(\R^{3})$, passing at $t=0$ through the point $\varphi_{{\bf a},{\bf p}}$, and such that  (\ref{pproj}) and (\ref{w}) are satisfied.
Suppose further that the evolution $\varphi=\varphi_{t}$ is unitary, so that, by Stone's theorem, $\frac{d\varphi}{dt}=-\frac{i}{\hbar} {\widehat H}\varphi$ for some self-adjoint operator ${\widehat H}$.
It is claimed then that the operator ${\widehat H}$ is uniquely defined and is equal to $-\frac{\hbar^{2}}{2m}\Delta+V({\bf x})$. 
In other words, the Schr{\"o}dinger evolution is the only unitary evolution on $L_{2}(\R^3)$ for which the system constrained to the classical phase space $M^{\sigma}_{3,3}$ satisfies Newtonian equations of motion for the particle.
\end{em}
\end{quote}
To prove, let us first verify that (\ref{pproj}) and (\ref{w}) imply the Ehrenfest theorem on states $\varphi \in M^{\sigma}_{3,3}$. As discussed, the Ehrenfest theorem can be written in the following form:
\begin{equation}
\label{EE1}
2\mathrm{Re} \left(\frac{d\varphi}{dt}, {\widehat x} \varphi \right)=\left(\varphi, \frac{\widehat p}{m}\varphi \right)
\end{equation}
and
\begin{equation}
\label{EE2}
2\mathrm{Re} \left(\frac{d\varphi}{dt}, {\widehat p} \varphi \right)=\left(\varphi, - \nabla V({\bf x}) \varphi \right).
\end{equation}
From (\ref{pproj}) and (\ref{del1}) we have at $t=0$,
\begin{equation}
\label{pproj2}
\frac{v^{\alpha}}{2\sigma}=\mathrm{Re}\left(\frac{d \varphi}{dt}, -\widehat{ \frac{\partial r}{\partial x^{\alpha}}}e^{i\theta}\right)=\frac{1}{\sigma}\mathrm{Re}\left(\frac{d\varphi}{dt}, (x-a)^{\alpha}\varphi\right).
\end{equation}
Because of the unitary condition, we have $\mathrm{Re}\left(\frac{d\varphi}{dt}, \varphi\right)=0$ and so (\ref{pproj2}) yields
\begin{equation}
\label{pproj3}
2\mathrm{Re}\left(\frac{d\varphi}{dt}, x^{\alpha}\varphi\right)=v^{\alpha}=\frac{p^{\alpha}}{m}.
\end{equation}
Together with 
$
\left(\varphi, {\widehat p}\varphi\right)=\left(\varphi, {\bf p}\varphi\right)={\bf p}
$
this gives the first Ehrenfest theorem (\ref{EE1}) on states $\varphi \in M^{\sigma}_{3,3}$.

Similarly, from (\ref{w}), (\ref{A}) and (\ref{del1}) we have at $t=0$,
\begin{equation}
\label{w2}
\frac{mw^{\alpha} \sigma}{\hbar}=\mathrm{Re} \left(\frac{d\varphi}{dt}, i\widehat{\frac{\partial\theta}{\partial p^{\alpha}}}\varphi\right)=\frac{\hbar}{\sigma}\mathrm{Re}\left(\frac{d\varphi}{dt}, \frac{i(x-a)^{\alpha}}{\hbar}\varphi\right),
\end{equation}
with
\begin{equation}
\label{WV}
mw^{\alpha}=-\left.\frac{\partial V({\bf x})}{\partial x^{\alpha}}\right|_{{\bf x}={\bf a}}.
\end{equation}
On the other hand, 
\begin{equation}
{\widehat p}\varphi=-i\hbar \nabla \varphi=-i\hbar\left(-\frac{{\bf x}-{\bf a}}{2\sigma^{2}}+\frac{i{\bf p}}{\hbar}\right)\varphi.
\end{equation}
Again, from the unitary condition, we have $\mathrm{Re}\left(\frac{d\varphi}{dt}, \varphi\right)=0$ and so we can rewrite (\ref{w2}) as 
\begin{equation}
\label{w3}
\frac{mw^{\alpha} \sigma}{\hbar}=\frac{\sigma}{\hbar}\mathrm{Re}\left(\frac{d\varphi}{dt}, {\widehat p}^{\alpha}\varphi \right),
\end{equation}
or,
\begin{equation}
\label{w4}
2\mathrm{Re}\left(\frac{d\varphi}{dt}, {\widehat p}^{\alpha}\varphi \right)=mw^{\alpha}.
\end{equation}
From this and (\ref{WV}), we get the second Ehrenfest theorem (\ref{EE2}) on states $\varphi \in M^{\sigma}_{3,3}$. 
Note that the components (\ref{pproj2}) and (\ref{w2}) are the real and imaginary parts of the classical phase space component of the velocity of state. In particular, the classical phase space submanifold inherits a complex structure from $CP^{L_{2}}$.

Now, from the derived Ehrenfest theorem and the Stone's theorem for a unitary evolution
\begin{equation}
\label{back1}
\frac{d\varphi}{dt}=-\frac{i}{\hbar} {\widehat H}\varphi,
\end{equation}
we get the following equations for the unknown self-adjoint operator ${\widehat H}$, valid for all functions $\varphi$ in $M^{\sigma}_{3,3}$:
\begin{equation}
\label{a}
\left( \varphi, i[{\widehat H},{\widehat x}]\varphi \right)=\frac{\hbar }{m}\left(\varphi, {\widehat p}\varphi\right)
\end{equation}
and
\begin{equation}
\label{b}
\left( \varphi, i[{\widehat H},{\widehat p}]\varphi \right)=\hbar\left(\varphi, -\nabla V({\bf x})\varphi\right).
\end{equation} 
Because $M^{\sigma}_{3,3}$ is complete in $L_{2}(\R^3)$, there exists a unique linear extension of the operators ${\widehat x}$, ${\widehat p}$ and $-\nabla V({\bf x})$  from $M^{\sigma}_{3,3}$ onto (a dense subset of) $L_{2}(\R^3)$. Likewise, for a given operator ${\widehat H}$, there exists a unique extension of the quadratic forms in the equations (\ref{a}) and (\ref{b}) from $M^{\sigma}_{3,3}$ to (a dense subset of) $L_{2}(\R^3)$. 
The resulting equations define ${\widehat H}$ uniquely. That is, there exists a unique operator ${\widehat H}$ for  which
\begin{equation}
\label{a1}
\left( f, i[{\widehat H},{\widehat x}]f \right)=\frac{\hbar }{m}\left(f, {\widehat p}f\right)
\end{equation}
and
\begin{equation}
\label{b1}
\left( f, i[{\widehat H},{\widehat p}]f \right)=\hbar\left(f, -\nabla V({\bf x})f\right)
\end{equation}
for all functions $f$ in the dense subset $D$ of $L_{2}(\R^{3})$, which is the common domain of all involved operators.
In fact, by choosing an orthonormal basis $\{e_{j}\}$ in $D$ and considering (\ref{a1}), (\ref{b1}) on functions $f=e_{k}+e_{l}$ and $f=e_{k}+ie_{l}$ we conclude that all matrix elements of the operators on the left and right of the equations (\ref{a1}) and  (\ref{b1}) must be equal. So the equations can be written in the operator form
\begin{equation}
\label{1}
i[{\widehat H},{\widehat x}]=\frac{\hbar }{m}{\widehat p}
\end{equation}
and
\begin{equation}
\label{2}
i[{\widehat H},{\widehat p}]=-\hbar\nabla V({\bf x}).
\end{equation}
From (\ref{1}) and (\ref{2}), it then follows that, up to an irrelevant constant, ${\widehat H}=\frac{{\widehat p}^{2}}{2m}+V({\bf x})$. 

Because (\ref{pproj}) and (\ref{w}) remain true for the potentials that depend on time and the equations used to obtain the result were considered at a fixed moment of time, the derivation remains valid for the time-dependent potentials $V({\bf x}, t)$ as well. Generalization to the case of $n$ interacting distinguishable particles described by tensor product of states (\ref{del1}) is straightforward and leads to the Hamiltonian 
${\widehat H}=\sum_{k}\frac{{\widehat p_{k}}^{2}}{2m_{k}}+V({\bf x}_{1}, ..., {\bf x}_{n})$.

By (\ref{omega}), a point $\varphi_{{\bf a},{\bf p}}$ in the classical phase space  $M^{\sigma}_{3,3}$ defines the initial position and velocity of the particle in $\R^3$. The solution of Newton's equations with this initial condition defines a unique classical path $({\bf a}_{t}, {\bf p}_{t})$ of the particle. Let's call the (non-linear) operator $U_{c}(t,0): M^{\sigma}_{3,3} \longrightarrow M^{\sigma}_{3,3}$, given by  
\begin{equation}
U_{c}(t,0)\left( \Omega({\bf a}_0, {\bf p}_0)\right)=\Omega({\bf a}_t, {\bf p}_t)
\end{equation}
with $\Omega$ given by (\ref{omega}), the {\it Newtonian evolution operator}. It was shown that there exists a unique unitary evolution operator  $U_{q}(t,0): L_{2}(\R^3) \longrightarrow L_{2}(\R^3)$, 
such that  $U_{q}(t,0)\varphi_{{\bf a},{\bf p}}=\varphi_{t}$ satisfies (\ref{pproj}) and (\ref{w}) for all $\varphi_{{\bf a},{\bf p}} \in M^{\sigma}_{3,3}$. 
It turned out to be the usual Schr{\"o}dinger evolution operator. The domain $L_{2}(\R^3)$ of this operator is the (closure of the) linear envelop of the domain $M^{\sigma}_{3,3}$ of the Newtonian evolution operator. The component of the velocity vector field $\frac{d U_{q}(t,0)\varphi_{{\bf a},{\bf p}}}{dt}$ tangent to $M^{\sigma}_{3,3}$ gives the usual Newtonian velocity and acceleration of the particle. The meaning of the additional components of $\frac{d \varphi}{dt}$ was revealed in (\ref{decomposition}).

 The obtained embedding of the classical phase space  into the space of states complemented by existence and uniqueness of extension of Newtonian to Schr{\"o}dinger evolution signifies that Newtonian dynamics found its full-fledged realization within the realm of quantum physics governed by the Schr{\"o}dinger equation. This realization is valid independently of whether it is taken to mean the actual physical embedding or only as a mathematical representation.

 \section{The Born rule and the normal probability distribution}
 \label{Born}

The isometric embedding of the classical space $M^{\sigma}_{3}$ into the space of states $L_{2}(\R^3)$ results in a relationship between distances in $\R^3$ and in the projective space $CP^{L_{2}}$.
The distance between two points ${\bf a}$ and  ${\bf b}$ in $\R^{3}$ is $\left\|{\bf a}-{\bf b}\right\|_{\R^{3}}$. Under the embedding of the classical space into the space of states, the variable ${\bf a}$ is represented by the state $\tilde{\delta}^{3}_{\bf a}$. The set of states $\tilde{\delta}^{3}_{\bf a}$ form a submanifold $M^{\sigma}_{3}$ in the Hilbert spaces of states $L_{2}(\R^{3})$, which is "twisted" in $L_{2}(\R^{3})$. It belongs to the sphere $S^{L_{2}}$ and spans all dimensions of $L_{2}(\R^{3})$. The distance between the states $\tilde{\delta}^{3}_{\bf a}$, $\tilde{\delta}^{3}_{\bf b}$ on the sphere $S^{L_{2}}$ or in the projective space $CP^{L_{2}}$ is not equal to $\left\|{\bf a}-{\bf b}\right\|_{\R^{3}}$. In fact, the former distance measures length of a geodesic between the states while the latter is obtained using the same metric on the space of states, but applied along a geodesic in the twisted manifold $M^{\sigma}_{3}$. 
The precise relation between the two distances is given by
\begin{equation}
\label{mainO}
e^{-\frac{({\bf a}-{\bf b})^{2}}{4\sigma^{2}}}=\cos^{2}\theta(\tilde{\delta}^{3}_{\bf a}, \tilde{\delta}^{3}_{\bf b}),
\end{equation}
where $\theta$ is the Fubini-Study distance between states in $CP^{L_{2}}$. The distance $\theta$ in the projective space of states $CP^{L_{2}}$ appears here for a good reason: the fibres of the fibration $\pi: S^{L_{2}} \longrightarrow CP^{L_{2}}$ through the points of the classical space $M^{\sigma}_{3}$ are orthogonal to this space. This is why the distance in  $M^{\sigma}_{3}$ can be expressed in terms of the distance in $CP^{L_{2}}$. Despite the non-trivial geometry contained in (\ref{mainO}), the formula itself is easy to derive. The left hand side is the result of integration in $|(\tilde{\delta^{3}_{\bf a}}, \tilde{\delta^{3}_{\bf b}})|^{2}$. On the other hand, the same expression is equal to the right side of (\ref{mainO}) by definition of the Fubini-Study metric.

The relation (\ref{mainO}) has an immediate implication onto the form of probability distributions of random variables over $M^{\sigma}_{3}$ and $CP^{L_{2}}$. 
In particular, consider a random variable $\psi$ over $CP^{L_{2}}$. Suppose that the restricted random variable  ${\widetilde \delta^{3}_{\bf a}}$, equivalently, ${\bf a}$, defined over $M^{\sigma}_{3}=\R^3$ is distributed normally on $\R^3$. Then the direction-independent probability distribution of $\psi$ satisfies the Born rule for the probability of transition between arbitrary states. The opposite is also true.
In other words, we claim that: 

\begin{quote} 
\begin{em}
The normal distribution law on $M^{\sigma}_{3}$ implies the Born rule on $CP^{L_{2}}$. 
Conversely,
the Born rule on the space of states implies the normal distribution law on $M^{\sigma}_{3}$.
\end{em}
\end{quote}
The fact that the Born rule implies the normal distribution on $M^{\sigma}_{3}$ is straightforward. According to the Born rule, the probability density $f({\bf b})$ to find the particle in a state ${\widetilde \delta^{3}_{\bf a}}$ at a point ${\bf b}$  is equal to
\begin{equation}
\label{Born1}
|{\widetilde \delta^{3}_{\bf a}}({\bf b})|^{2}=|({\widetilde \delta^{3}_{\bf a}}, \delta^{3}_{\bf b})|^{2}=\left(\frac{1}{2\pi \sigma^{2}}\right)^{3/2}e^{-\frac{({\bf a}-{\bf b})^{2}}{2\sigma^{2}}}\equiv f_{{\bf a}, \sigma}({\bf b}),
\end{equation}
which is the normal distribution function. It follows that on the elements of $M^{\sigma}_{3}$, the Born rule {\it is} the rule of normal distribution.

The Born rule on $M^{\sigma}_{3}$ can be also written in term of the probability $P(\tilde{\delta}^{3}_{\bf a}, \tilde{\delta}^{3}_{\bf b})$ of transition between the states $\tilde{\delta}^{3}_{\bf a}, \tilde{\delta}^{3}_{\bf b}$ in $M^{\sigma}_{3}$:
\begin{equation}
\label{Born2}
P(\tilde{\delta^{3}_{\bf a}}, \tilde{\delta^{3}_{\bf b}})=|(\tilde{\delta^{3}_{\bf a}}, \tilde{\delta^{3}_{\bf b}})|^{2}.
\end{equation}
Assuming ${\widetilde \delta^{3}}_{\bf b}$ is sufficiently sharp, the formulas (\ref{Born1}) and (\ref{Born2}) mean the same thing. 
In fact,
\begin{equation}
\label{Born4}
|({\widetilde \delta^{3}}_{\bf a}, {\widetilde \delta^{3}}_{\bf b})|^{2} =  f_{{\bf a}, \sqrt{2}\sigma}({\bf b})(\Delta x)^{3},
\end{equation}
where $f_{{\bf a}, \sqrt{2}\sigma}$  is the normal distribution function with standard deviation ${\sqrt 2}\sigma$ and $\Delta x=\sqrt{4\pi\sigma^2}$. This relates the probability in (\ref{Born2}) to the normal probability density in (\ref{Born1}) and identifies $P(\tilde{\delta^{3}_{\bf a}}, \tilde{\delta^{3}_{\bf b}})$ with the probability of finding the macroscopic particle near the point ${\bf b}$.

Conversely, suppose we have a rule for probability of transition between states in $CP^{L_{2}}$ which gives the normal distribution law for the states in $M^{\sigma}_{3}$ and depends only on the distance between states. Let's show that this must be the Born rule.
In fact, the Fubini-Study distance between the states $\tilde{\delta^{3}_{\bf a}}$, $\tilde{\delta^{3}_{\bf b}}$ takes on all values from $0$ to $\pi/2$, which is the largest possible distance between points in $CP^{L_{2}}$. By assumption, the probability $P(\varphi, \psi)$ of transition between any states $\varphi$ and $\psi$ depends only on the Fubini-Study distance 
$\theta(\pi(\varphi), \pi(\psi))$ between the states. Given arbitrary states 
$\varphi, \psi \in S^{L_{2}}$, let then $\tilde{\delta^{3}_{\bf a}}$,
$\tilde{\delta^{3}_{\bf b}}$ be two states in $M^{3}_{\sigma}$, such that
\begin{equation}
\theta(\pi(\varphi), \pi(\psi))=\theta(\tilde{\delta^{3}_{\bf a}}, 
\tilde{\delta^{3}_{\bf b}}).
\end{equation}
It then follows that
\begin{equation}
P(\varphi, \psi)=P(\tilde{\delta^{3}_{\bf a}}, \tilde{\delta^{3}_{\bf b}})=\cos^{2}\theta(\tilde{\delta^{3}_{\bf a}}, \tilde{\delta^{3}_{\bf b}})
=\cos^{2}\theta(\pi(\varphi), \pi(\psi)),
\end{equation} 
which yields the Born rule for arbitrary states and proves the claim.

\section{The Born rule for a measurement of spin}
\label{BornSpin}

We are now in a position to compare the process of measurement in the classical and quantum physics. 
First of all, the classical space and phase space are now submanifolds in the Hilbert space of states. This allows us to use the same language when analyzing both types of measurement. Second, the Newtonian dynamics is now a restriction of the Schr{\"o}dinger dynamics to the classical phase space submanifold. Conversely, the Schr{\"o}dinger dynamics is a unique unitary extension of the Newtonian dynamics from the classical phase space to the Hilbert space. This allows us to begin with a model of measurement satisfying Newton laws and extend it to a model consistent with the rules of quantum mechanics. Finally, the normal probability law is the restriction of the Born rule to the classical space submanifold. Conversely, the Born rule is the unique isotropic extension of the normal probability law from the classical space to the space of states. In particular, a classical model of measurement with a normal distribution of the measured quantity should lead us to a model consistent with the Schr{\"o}dinger dynamics and the Born rule.

Measurements performed on a macroscopic particle satisfy generically the normal distribution law for the measured observable. This is consistent with the central limit theorem and indicates that the specific way in which the observable was measured is not important. 
For example, consider measurements of position of a particle. 
A common way of finding the position of a macroscopic particle is to expose it to light of sufficiently short wavelength and to observe the scattered photons. Due to the unknown path of the incident photons, multiple scattering events on the particle, random change in position of the particle, etc., the process of observation can be described by the diffusion equation with the observed position of the particle experiencing Brownian motion from an initial point during the time of observation. This results in the normal distribution of observed position of the particle.

The ability to describe 
measurements on a macroscopic particle as a diffusion seems to be a general feature of measurements in the macro-world, independent of a particular measurement set-up. The averaging process making the central limit theorem applicable and leading to the normal distribution of the position random variable can be seen, for example, as the result of random hits experienced by the particle from the surrounding particles participating in the measurement. These random hits are equally likely to come from any direction, independent of the initial position of the particle, leading to Brownian motion and the validity of the diffusion equation for the probability density of the position random variable for the particle. 

It is claimed now that at any time $t$, the initial state of a microscopic particle undergoing a measurement
is equally likely to shift in any direction in the tangent space to the appropriate projective space of states. In this and the following sections it will be demonstrated that this fact may be responsible for the validity of the Born rule for the probability of transition of quantum states. 
As a model example, in this section we will derive the Born rule for a measurement of the spin state of a non-relativistic electron. 

For this, let us return to the space $\C^2$ of spin $1/2$ particle considered in section \ref{curvature} and consider the Pauli equation for the electron interacting with a spin-measuring device.  Let us assume that under the measurement the Stern-Gerlach interaction term in the equation drives the system, so that other terms can be neglected. (A discussion of when the interaction term can be assumed to drive the system is given in section \ref{measurements}.) In this case the Hamiltonian of interaction between the electron and the device is given by  ${\widehat h}=-\mu {\widehat {\bf \sigma}} \cdot {\bf B}$, where ${\bf B}$ is the magnetic field, ${\bf {\widehat \sigma}}=({\widehat \sigma}_x, {\widehat \sigma}_{y}, {\widehat \sigma}_{z})$ and $\mu=e/2m$.
The evolution equation for the spin states in the space $\C^2$ is then given by
\begin{equation}
\label{spin_ev}
i\hbar \frac{d \varphi_{t}}{dt}=-\mu {\widehat {\bf \sigma}} \cdot {\bf B}\varphi_{t}.
\end{equation}
Using 
\begin{equation}
\label{sigma0}
({\bf {\widehat \sigma}} \cdot {\bf A})({\bf {\widehat \sigma}} \cdot {\bf B})={\bf A}\cdot {\bf B}+i{\bf {\widehat \sigma}} \cdot {\bf A} \times {\bf B},
\end{equation} 
we obtain
\begin{equation}
\label{sigma1}
({\bf {\widehat \sigma}} \cdot {\bf B})^{2}={\bf B}^{2}.
\end{equation}
Since the matrix ${\widehat {\bf \sigma}} \cdot {\bf B}$ is hermitian, we have then 
\begin{equation}
\label{speed}
\left\| \frac{d \varphi_{t}}{dt} \right \|_{\C^{2}}=\left(\frac{i}{\hbar}\mu {\widehat {\bf \sigma}} \cdot {\bf B}\varphi_{t}, \frac{i}{\hbar}\mu {\widehat {\bf \sigma}} \cdot {\bf B}\varphi_{t}\right)^{\frac{1}{2}}_{\C^{2}}=\frac{\mu B}{\hbar},
\end{equation}
where $B$ is the norm of ${\bf B}$. 

Suppose the $z$-component of spin of an electron in a superposition of eigenstates of ${\widehat \sigma}_z$ is measured. For instance, we could insert a screen behind a Stern-Gerlach magnet and observe where the electron lands on the screen. When the electron interacts with the screen, it experiences a random magnetic field created by the molecules and atoms of the screen and their thermal motion. Because of the physical symmetry and the central limit theorem, the components of the resulting magnetic field ${\bf B}$ can be assumed to be independent, identically distributed, normal random variables. In this case the vector  $i{\widehat {\bf \sigma}} \cdot {\bf B}$ in the Lie algebra $su(2)$ with the Killing form is a normal random vector with an isotropic probability distribution, so that the level surfaces of the probability density are spheres. In particular, from (\ref{spin_ev}) we conclude that any direction of the displacement $\delta \varphi$ of the initial spin state $\varphi_{0}$ in the tangent space $T_{\varphi_{0}}S^3$ to the sphere of states $S^3$ with the Killing metric is equally likely. Also, the distribution of the displacements is the same for all initial states $\varphi_{0}$, i.e., in all tangent spaces $T_{\varphi_{0}}S^3$, and is normal.

Let us look at the resulting motion of state in the projective space $CP^1$ of physical states. 
For this consider the complex lines $\{\varphi\}$ formed for each state
$\varphi=\left[ \begin{array}{c}
\varphi_{1} \\ 
\varphi_{2}
\end{array}
\right]$
 by the vectors $\lambda \varphi$, $\lambda \in \C$.
Provided $\varphi_{1} \neq 0$, there is a unique point of intersection of the line $\{\varphi\}$ with the affine plane of vectors
$\left[ \begin{array}{c}
1 \\ 
\xi
\end{array}
\right], \xi \in \C$ in $\C^2$. Namely, by setting 
\begin{equation}
\lambda \left[ \begin{array}{c}
\varphi_{1} \\ 
\varphi_{2}
\end{array}
\right]=\left[ \begin{array}{c}
1 \\ 
\xi
\end{array}
\right],
\end{equation}
we obtain 
\begin{equation}
\label{xi}
\xi=\frac{\varphi_{2}}{\varphi_{1}}, 
\end{equation}
and $\lambda=1/\varphi_1$. The map $u=\{\varphi\} \longrightarrow \xi$ provides a coordinate chart on $CP^{1}$ that identifies $CP^{1}$ excluding a point (the complex line through
$\left[ \begin{array}{c}
0 \\ 
1
\end{array}
\right]$)
with the set $\C$ of complex numbers. 
Under the isomorphism $\widehat{\omega}$ in (\ref{MatN}), the vectors
$\left[ \begin{array}{c}
1 \\ 
\xi
\end{array}
\right]$ form an affine subspace in the Lie algebra $su(2)$. The algebra $su(2)$ with the Killing form is the Euclidean space $\R^{3}$ of vectors ${\bf x}= \sum_{k} x^{k} i{\widehat \sigma}_{k}$. The stereographic projection then identifies the unit sphere $S^{2}$ at the origin of $\R^{3}$ with the above plane $\C$ plus a point, i.e., with $CP^{1}$ itself. Moreover, the usual metric on $S^2$ induced by its embedding into $\R^3$ is the Fubini-Study metric on $CP^1$.
The relationship of the coordinate $\xi$ in the plane $\C$ with coordinates $(x^{1}=x,x^{2}=y,x^{3}=z)$ of the corresponding point on the sphere $S^{2}$ is given by
\begin{equation}
\xi=\frac{x+iy}{1-z}.
\end{equation}
Solving this for $x,y$ and $z$ and using (\ref{xi}), one obtains:
\begin{eqnarray}
\label{xyz}
x&=&\varphi_{1}{\overline \varphi_{2}}+{\overline \varphi_{1}}\varphi_{2},\\
\label{xyz1}
y&=&i(\varphi_{1}{\overline \varphi_{2}}-{\overline \varphi_{1}}\varphi_{2}), \\
\label{xyz2}
z&=&\varphi_{2}{\overline \varphi_{2}}-\varphi_{1}{\overline \varphi_{1}}.
\end{eqnarray}
The resulting map $\pi: S^{3} \longrightarrow S^{2}$ given by $\pi(\varphi_{1},\varphi_{2}) = (x,y,z)$ is a coordinate form of the bundle projection from the sphere of unit states in $\C^2$ onto the space $CP^1$ of physical states. The map $\pi$ relates the spaces of representation of the groups $SU(2)$ and $SO(3)$ and maps the spin state of a particle to a vector in $\R^3$.

By writing $\varphi_1=e^{i \alpha_1}\cos \eta$ and $\varphi_2=e^{i \alpha_2}\sin \eta$, where $\eta \in [0,\frac{\pi}{2}]$ and $\alpha_1, \alpha_2 \in [0,2\pi)$,  and
\begin{equation}
\lambda \left[ \begin{array}{c}
\varphi_{1} \\ 
\varphi_{2}
\end{array}
\right]=
\left[ \begin{array}{c}
\cos \eta \\ 
e^{i (\alpha_2 - \alpha_1)}\sin \eta
\end{array}
\right]
\end{equation}
with $\lambda=e^{-i \alpha_1}$, we obtain $\xi=e^{i(\alpha_2-\alpha_1)}\tan \eta$ for the variable $\xi$ in (\ref{xi}). 
The resulting Hopf coordinates $(\eta, \alpha_1,\alpha_2)$ are particularly useful in visualizing the fibration $\pi: S^{3} \longrightarrow S^{2}$. Let $(\theta, \phi)$ be the usual spherical coordinates on $S^2$. We then have $\theta=\pi-2\eta$ and $\phi=\alpha_2-\alpha_1$ so that $\xi=e^{i\phi}\cot \frac{\theta}{2}$. By fixing $\eta$ and letting $\alpha_2-\alpha_1$ run, we obtain a parallel (fixed $\theta$, varying $\phi$) on $S^2$. On the other hand, fixing $\eta$ and $\alpha_2-\alpha_1$ and varying $\alpha_1+\alpha_2$, we specify the state $\{\varphi\} \in CP^1$ while traveling around the fibre (the phase circle $S^1$). For a fixed $\eta$, the coordinates  $\alpha_2-\alpha_1$ and $\alpha_1+\alpha_2$ parametrize a torus. The volume form on $S^3$ is given by 
\begin{equation}
\label{volumeS}
dV=\sin \eta \cos \eta d\eta \wedge d \alpha_1 \wedge d \alpha_2=\frac{1}{8} \sin \theta d\theta \wedge d\phi \wedge d \alpha=\frac{1}{8} dA \wedge d \alpha,
\end{equation} 
where $\alpha=\alpha_1+\alpha_2$ and $dA$ is the volume form on the sphere $S^2$.

Let us now return to the motion of spin-state $\varphi \in S^3$ under a measurement of the $z$-component of spin. It was argued that at the moment of observation such a motion can be described by the equation (\ref{spin_ev}) with the normal random magnetic field ${\bf B}$. The state $\varphi$ performs then a random walk on the sphere $S^3$.
Since the initial state $\varphi_0$ is defined only up to a phase factor, we are dealing with an ensemble of states with uniformly distributed phases. Furthermore, since the evolution equation is linear, a constant initial phase factor is preserved throughout the evolution. The random walk of the state $\varphi$ on $S^3$ can be then described in terms of a random walk of the physical state $\{\varphi\}$ on the sphere $S^2=CP^1$. The restriction of the volume form $dV$ in (\ref{volumeS}) yields the usual area form $dA=\sin \theta d\theta\wedge d\varphi$ on $S^2$. The distribution of the increments $\delta \{\varphi\}$ in the tangent space $T_{\{\varphi_{0}\}}S^2$ for any initial state $\{\varphi_0\}=(\theta_0,\phi_0)$ is normal. Therefore, during a sufficiently short time step $\tau$ the density of states and therefore the probability density function for the state is nearly Gaussian in the variables $\theta-\theta_0, \phi-\phi_0$. Disregarding the change in the values of $\phi$, the marginal probability of a random move in $\theta$ is then proportional to $\rho(\theta-\theta_0) \sin \theta d\theta$, where $\rho$ is Gaussian.  Because $z=\cos \theta$, we have $\sin \theta d\theta=-dz$ and so the probability of steps of an equal small increment $dz$ is approximately the same for each step, independent of the value of $z$. The process can be then modeled by a simple symmetric random walk on the $z$-axis. 

Let then $dz=\pm \Delta$ be the step of the walk with $\Delta \ll 1$ and the positive and negative values being equally likely. 
From (\ref{xyz2}) it follows that 
\begin{equation}
\label{amp}
|\varphi_{1}|^2  =  \frac{1+z}{2}, \  \textrm{and}  \ \ |\varphi_{2}|^2 =   \frac{1-z}{2}.
\end{equation}
The gambler's ruin mechanism tells us now that the probability $P_{2}$ for the state $\varphi$ to reach the state $\left[ \begin{array}{c}
0 \\ 
1
\end{array}
\right]$ ($z=-1$) first, as a result of the described random motion is  equal to 
\begin{equation}
P_{2} \ = \ \frac{ \textrm{number of steps from} \ z \  \textrm{to}  \ -1}{\textrm{number of steps from} \ -1 \ \textrm{to} \ 1} \ = \ \frac{1-z}{2} \ = \ |\varphi_{2}|^2.
\end{equation} 
Similarly, the probability $P_{1}$ that $\varphi$ will reach the state 
$\left[ \begin{array}{c}
1 \\ 
0
\end{array}
\right]$
first is given by 
$P_{1}=|\varphi_{1}|^2$. This is the Born rule for transitions of spin-states.

Let us review the key parts of this derivation. First, the random walk takes place on the space of states and is described by the Schr{\"o}dinger (Pauli) equation with a random, normally distributed magnetic field. Because of the group symmetry, the distribution of the increments of state under the random walk does not depend on a particular initial point $\varphi_0 \in S^3$ (homogenuity)  or the direction in the tangent space $T_{\varphi_{0}}S^3$ (isotropy). The metric on $S^3$ yields the Fubini-Study metric on $CP^1=S^2$ and the volume form on $S^3$ induces the usual area form on $S^2$. The probability of steps $dz$ towards the eigenstates of ${\widehat \sigma}_{z}$ in the random walk on the $z$-axis turns out to be approximately the same for small, equally sized steps. The coefficients in the decomposition of the state in the basis of eigenvectors determine how far the state needs to go to reach a particular eigenstate and the probability of that event. The derivation applies to a measurement of any component of spin. If upon reaching one of the eigenstates the state is absorbed (say, the electron in the upper arm of the Stern-Gerlach magnet is absorbed by the screen), the process stops.

\section{The Born rule for a measurement of position}
\label{measurements}

In section \ref{Born}, the Born rule for transition of states in the space $CP^{L_{2}}$ was derived from the normal distribution law on the submanifold $M^{\sigma}_3$. The derivation used an additional assumption that the probability of transitions depends only on the distance between states. In section \ref{BornSpin}, the evolution of spin-state of an electron in a random, normally distributed magnetic field was considered.
The probability distribution of the displacement vector $\delta \varphi$ of the spin-state was shown to be independent of the initial state $\varphi_0$ and the direction of $\delta \varphi$ in the tangent space $T_{\{\varphi_0\}}CP^{1}$. In other words, the distribution of states driven by the field may depend only on the Fubini-Study distance between the states. 
It will be now argued that a measurement of the position of a microscopic particle yields similar results. That is, the probability distribution of the state random variable may depend only on the Fubini-Study distance between states. The Born rule for transition of states of the particle then follows.

To be specific, consider a particle exposed to a stream of photons of sufficiently high frequency and number density. The scattered photons are then observed to determine the position of the particle. The field of photons in the experiment will be treated classically, as a fluctuating potential in a region surrounding the source. Despite the classical treatment of the field and other assumptions made about the potential, a more general proof in section \ref{continuity} will confirm that the result derived here is general.

Recall first that the space $M^{\sigma}_{3,3}$  is complete in $L_{2}(\R^3)$. Consider the subset of $M^{\sigma}_{3,3}$ formed by the states
\begin{equation}
\label{initial1}
    \varphi_{{\bf m}{\bf n}}({\bf x})=\left(\frac{1}{2\pi\sigma^{2}}\right)^{3/4}e^{-\frac{({\bf x}-\alpha{\bf n})^{2}}{4\sigma^{2}}}e^{i\frac{\beta{\bf m} {\bf x}}{\hbar}},
\end{equation}
where $\alpha=\sqrt{2\pi} \sigma$,  $\beta=\frac{h}{\sqrt{2\pi} \sigma}$  and ${\bf m}, {\bf n}$ take values on the lattice ${\Z}^{3} \times {\Z}^{3}$ of points with integer coordinates in $\R^3 \times \R^3$. The set of functions (\ref{initial1}) is known to be also complete in $L_{2}(\R^3)$. Any state in $L_{2}(\R^3)$ can be then represented by a linear combination of states $\varphi_{{\bf m}{\bf n}}$. (For $\alpha \beta <h$ the system of functions $ \varphi_{{\bf m}{\bf n}}$ is called the Gabor or Weil-Heisenberg frame.) 
In particular, the initial state $\psi$ of the particle can be represented by a sum
\begin{equation}
\label{supp}
\psi=\sum_{{\bf m},{\bf n}} C_{{\bf m}{\bf n}}\varphi_{{\bf m}{\bf n}}.
\end{equation}

The set $M^{\sigma}_{3}$ is also complete in $L_{2}(\R^3)$. Here too there exist countable subsets of $M^{\sigma}_{3}$ that are complete in $L_{2}(\R^3)$. Moreover, an arbitrary initial state $\psi$ in $L_{2}(\R^3)$ can be approximated as well as necessary by a finite discrete sum
\begin{equation}
\label{suppN}
\psi \approx \sum_{\bf n} C_{\bf n} {\widetilde \delta}^{3}_{{\bf a}-\gamma{\bf n}},
\end{equation}
where ${\bf a}$ is arbitrary, ${\bf n} \in \Z^3$, and the value of $\gamma>0$ together with the number of terms in the sum depend on $\psi$ and the needed approximation. The same is true when the Gaussian functions in (\ref{suppN}) are replaced with the indicator functions. Namely, taking $\gamma$ sufficiently small and partitioning the space $\R^3$ into the cubical cells of edge $\gamma$ centered at the lattice points  ${\bf a}-\gamma {\bf n}$, we have 
\begin{equation}
\label{appx}
\psi \approx \sum_{\bf n} C_{\bf n} \chi_{{\bf n}}.
\end{equation}
Here $\chi_{\bf n}$ is the unit-normalized indicator function of the ${\bf n}$th cell. The potential can be written as a sum $\sum_{{\bf n}} V_{\bf n}{\widehat P}_{{\bf n}}$, where ${\widehat P}_{\bf n}$ is the projector onto the subspace of functions with support in the ${\bf n}$th cell. The components $V_{\bf n}$ for different values  of ${\bf n}$ will be assumed to be independent, identically distributed, normal random variables. In the case of position measurement by scattering photons off the particle, the component $V_{\bf n}$ can be associated with a photon in the ${\bf n}$th cell at time $t$.

For simplicity, let us neglect the kinetic energy term in the Hamiltonian ${\widehat h}$. We will see when the resulting approximation is valid later.  Let us denote the solution of the Shr{\"o}dinger equation with the initial state $\psi$ by  $\Psi(t)$ and set   $\Psi(t)= e^{-\frac{i{\overline V}t}{\hbar}} \psi(t)$, where ${\overline V}=({\widehat V}\psi, \psi)$ and $\psi(0)=\psi$.
 We then have at $t=0$
 \begin{equation}
\label{VVn}
\frac{d \psi}{dt}=-\frac{i}{\hbar}{\widehat V}_{\perp}\psi,
\end{equation}
where ${\widehat V}_{\perp}={\widehat V}-{\overline V}I$, as before. This equation gives the velocity of the state $\Psi(t)$  in the projective space $CP^{L_{2}}$ at $t=0$. Note that the velocity is zero where the potential  ${\widehat V}_{\perp}$ vanishes. For simplicity, we will consider the case when the potential ${\widehat V}_{\perp}$ acts on a compact subset $D$ of $\R^3$. In this case, only the projection of the initial state $\psi$ onto $D$ will be relevant for the outcomes of a position measurement in $D$. In particular, we can assume that the support of $\psi$ is in $D$.

The set up is very similar to the one encountered in section \ref{BornSpin}. The random magnetic field ${\bf B}$ in $\R^3$ is now replaced with the $n$-component random vector ${\bf V}_{\perp}$ with components $V_{\perp n}=V_{n}-{\overline V}$. The generators $i{\widehat \sigma}\cdot {\bf B}$ of the Lie algebra $su(2)$ are replaced with the operators $A_V$ defined by $iA_V=\sum_{{\bf n}} V_{\bf n}{\widehat P}_{{\bf n}}-{\overline V}I$ in the Lie algebra of the unitary group $U(N)$, where $N$ is the number of cells in $D$. Although the operators $A_V$ do not span the Lie algebra of the group $U(N)$, the one-parameter subgroups of $U(N)$ generated by these operators sweep out the space of states $CP^{N-1}$. Furthermore, the Killing metric on $U(N)$ yields the Fubini-Study metric on $CP^{N-1}$. Under such a realization of the symmetric space $CP^{N-1}$, a particular initial state $\{\psi\}$ is identified with the identity of the group and the generators $A_V$ form the tangent space $T_{\{\psi\}}CP^{N-1}$. From the definition of ${\overline V}$ and the decomposition (\ref{appx}), we have
\begin{equation}
{\overline V}=\sum_{\bf n} V_{\bf n}|C_{\bf n}|^2.
\end{equation}
Because $\sum |C_{\bf n}|^2=1$, the mean value of the random variable $V_{\perp m}$  is zero:
\begin{equation}
E(V_{\bf m}-{\overline V})=E(V_{\bf m})-E(V_{\bf m})\sum_{\bf n} |C_{\bf n}|^2=0.
\end{equation}
So the components $V_{\perp m}$ are independent, identically distributed normal random variables with the zero mean. It follows that at $t=0$ all directions of the velocity vector $\frac{d \psi}{dt}=-\frac{i}{\hbar}{\widehat V}_{\perp}\psi$ in the tangent space $T_{\{\psi\}}CP^{N-1}$ are equally likely. Furthermore, the acton of the unitary group on $CP^{N-1}$ is transitive. By moving the initial state $\{\psi\}$ and the tangent space $T_{\{\psi\}}CP^{N-1}$ around, we conclude that the distribution of the velocity vector is also independent of the initial state $\{\psi\}$. 
It follows that under the evolution (\ref{VVn}), the distribution of states $\{\psi\}$ can depend only on the Fubini-Study distance between the initial and the end states.

Let us check that the assumptions used in the derivation of the isotropy of the distribution of the displacement random variable $\delta \psi$ are realistic. 
Suppose for example that the position of an electron is measured by subjecting it to a stream of photons. 
Assume first that the initial state of the electron belongs to the classical phase space submanifold $M^{\sigma}_{3,3}$ of the space of states. Suppose also that the wave length of the photons is of the order of $1nm=10^{-9}m$ (x-rays) or larger.  Let us estimate the terms of the decomposition (\ref{decomposition}) for the velocity of the state of the electron. From the Compton scattering formula, we have for the difference in wave length of the incoming and scattered photons
\begin{equation}
\lambda_{f}-\lambda_{i} = \frac{h}{mc}(1-\cos \theta) \sim 10^{-12}m.
\end{equation}
The transferred energy is then
\begin{equation}
\Delta E=\frac{hc}{\lambda_{i}}-\frac{hc}{\lambda_{f}} \sim 10^{-20}J.
\end{equation}
With the electron initially at rest, we have for the speed $v$ acquired during the interaction
\begin{equation}
\frac{mv^2}{2} \sim 10^{-20}J, \ \textrm{or}, \ v \sim 10^5m/s.
\end{equation}
The accuracy of position measurement is limited by the wave length. Setting $\sigma=\lambda \sim 10^{-9}m$, we have
for the classical velocity component of $\frac{d \varphi}{dt}$, given by the second term in (\ref{decomposition})
\begin{equation}
\frac{v}{2\sigma} \sim \frac{10^5}{10^{-9}}=10^{14}s^{-1}. 
\end{equation}
Estimating the time of interaction $\tau$ by $\lambda/c \sim 10^{-17}s$, we have
for the classical acceleration component, given by the third term in (\ref{decomposition}):
\begin{equation}
\frac{m w \sigma}{\hbar}=10^{17}s^{-1}.
\end{equation}
For the spreading component, given by the last term in (\ref{decomposition}), we obtain
\begin{equation}
\frac{\hbar}{4\sqrt{2}\sigma^2 m} \sim 10^{13}s^{-1}.
\end{equation}
In the estimate, the acceleration term is the largest of the three. 
The resolution parameter $\sigma$ in the non-relativistic position measurement experiments is typically much larger than the used value of $1nm$. With the increase in $\sigma$ (keeping $\lambda=\sigma$), the velocity term decreases as $\sigma^{-\frac{3}{2}}$, the acceleration term decreases as $\sigma^{-\frac{1}{2}}$ while the spreading term decreases as $\sigma^{-2}$. In particular, for the scattering of visible light we have $\lambda \sim 10^{-5}m$. This gives for the electron the velocity term of the order of $10^{8}s^{-1}$, the acceleration term $\sim 10^{15}s^{-1}$ and the spreading term $\sim 10^{5}s^{-1}$. Furthermore, if the mass $m$ increases, the value of the velocity term further decreases as $m^{-\frac{1}{2}}$, the value of the acceleration term increases as $m^{\frac{1}{2}}$, while  the spreading term decreases as $m^{-1}$, showing that the acceleration terms is by far the dominant term under these conditions.

Let us now write an arbitrary initial state $\psi$ as a superposition (\ref{supp}) of states in $M^{\sigma}_{3,3}$.  Then the variation $\delta \psi =\frac{d \psi}{dt}\tau$ can be also written as a series in functions from $M^{\sigma}_{3,3}$, so that each term of the series is a constant times a function in $M^{\sigma}_{3,3}$. The initial speed $v$ of each component function in $M^{\sigma}_{3,3}$ is limited by the speed of light $c$. If $v$ is of the same order as $c$, then the velocity and acceleration terms in the component function are of the same order. However, given the non-relativistic character of the problem, the major terms in the series correspond to $v \ll c$.  The spreading term in each term of the series is the same and is much smaller than the acceleration term. Therefore, given the near-orthogonality of the terms of the series, we can neglect the velocity and spreading parts in each term, which amounts to keeping only the potential term in the Hamiltonian. In particular, the motion of the state in these conditions amounts to a jiggling of the wave packet without much spreading or displacement.

Let us check now that under reasonably general  measurement conditions, the periods of a free evolution of the electron state can be neglected. In other words, interaction with the electromagnetic field is happening continuously in time. From the number density of photons, we can estimate the number of photons in one cubic meter of space by  $N \approx 2.02 \times 10^7 T^3$ and the average energy of a photon by $2.7 k_{B}T$, where $k_{B}$ is the Boltzmann constant and $T$ is temperature. For instance, taking $T \sim 500K$, we obtain $N \sim 10^{15}$. The photon with the average energy at this temperature has the wave length $\lambda \sim 10^{-5}m$. Under these conditions, at any time $t$ there is about one photon per cube of the volume $\lambda^{3}$.  So, at any $t$, each $M^{\sigma}_{3,3}$ component of $\psi$ experiences the potential of a photon passing by. Given these conditions, neglecting the free evolution of the electron state is a reasonable approximation.

Despite being heuristic, these estimates demonstrate that during the type of measurement considered in this section, the potential term is the main term in the Hamiltonian responsible for the dynamics of the particle under the Schr{\"o}dinger evolution.  On the other hand, if the number density of photons is significantly lower, the evolution will consist of a free Sch{\"o}dinger evolution combined with the periods when the state is driven by the potential term alone. We will return to this issue in section \ref{continuity}, where an alternative approach to the problem will be discussed.

When measuring the position of a macroscopic particle, the observed particle is exposed to a random potential that is responsible for the normal distribution of the position random variable. The motion of the particle can be in this case described by the Langevin equation. In the Hamiltonian description of interaction of a macroscopic particle with the surroundings (as in the Ullersma model \cite{Sanz}), the friction term comes from a contribution of the particles in the surroundings to the total potential in the Hamiltonian. In this sense the Langevin equation describes the Newtonian evolution of a system of particles.
On the other hand, in section \ref{components} it was verified that the Schr{\"o}dinger evolution with the state constrained to the manifold $M^{\sigma}_{3n,3n}$ yields the Newtonian evolution, and vice versa. So the Schr{\"o}dinger equation for the particle should be consistent with the Langevin equation.  For a hint of how this relationship may work, note that the random force term with the Gaussian distribution in the Langevin equation can be found in the potential term ${\widehat V}_{\perp}={\widehat V}-{\overline V}$ in (\ref{VVn}). Namely, for the states in $M^{\sigma}_{3,3}$ the term ${\widehat V}_{\perp}\psi$ yields the gradient of the function $V$ in a cell. 

The details of the relationship between the two equations will not be discussed in this paper. What is important here is that: (1) the distribution of the position random variable for a macroscopic particle undergoing a position measurement is Gaussian and (2) the distribution of states of a microscopic particle undergoing a similar measurement and exposed to the like-potential depends only on the Fubini-Study distance from the initial state in the projective space of states.
From  this and the derivation in the section \ref{Born}, it follows that the probability for the state of reaching a neighborhood of a point $\varphi$ is given by the Born rule: $P(\varphi,\psi)=|(\varphi, \psi)|^2$.

Given the lack of Lebesgue or any non-trivial translation-invariant measure on an infinite-dimensional Hilbert space, one may wonder how the state would have any chance of reaching a neighborhood of a given point in the case of an infinite-dimensional space of states $CP^{L_{2}}$. However, a realistic measuring device occupies a finite volume in the classical space. So the potential created by it can only affect a bounded region $D$ in $\R^3$. The initial state $\psi$ of the particle can be split onto the state $\psi_{D}$ that is the restriction of $\psi$ to $D$ and the leftover state $\zeta=\psi-\psi_{D}$. Because the action of the potential on $\zeta$ is trivial, the state $\zeta$ in the considered approximation is not going to change and will not participate in the measurement (the probability for it of reaching a detector in $D$ is zero). 
By (\ref{appx}), the state $\psi_{D}$ is a finite linear combination of the indicator functions $\chi_{\bf n}$ of the cells in $D$. Furthermore, under the motion in the random potential ${\widehat V}$ described by equation (\ref{VVn}), the state will continue to stay in  the finite-dimensional linear envelop  $L_{D}$ of the indicator functions of the cells in $D$.
In particular, the Lebesgue volume of a ball of a positive radius in $L_{D}$ exists and is positive. 
It follows that the state $\psi_{D}$ has a non-vanishing probability of reaching a neighborhood of the state ${\widetilde \delta}^{3}_{\bf a}$ and the relative probabilities of reaching neighborhoods of states ${\widetilde \delta}^{3}_{\bf a}$ for different points ${\bf a}$ are given by the Born rule.

\section{The motion of state under measurement}
\label{continuity}

Let us now look into details of the stochastic motion of a state under a measurement. Note that in the non-relativistic quantum mechanics, the particle, and therefore its state in a single particle Hilbert space, cannot disappear or get created.
The unitary property of evolution means that the state  can only move along the unit sphere in the space of states $L_{2}(\R^3)$.  To express this conservation of states in the case of observation of position of the particle, consider the density of states functional $\rho_{t} [\varphi;\psi]$. Here we begin with an ensemble of particles whose initial state belongs to a neighborhood of the state $\psi$ on the sphere of states $S^{L_{D}}\subset L_{D}$. The functional $\rho_{t} [\varphi;\psi]$ measures the number of states that by the time $t$ belong to a neighborhood of a state $\varphi \subset S^{L_{D}}$. It is approximately equal to the number of states in a small region around $\varphi$ in  $S^{L_{D}}$ divided by the volume of the region.
Under the isometric embedding $\omega: \R^3 \longrightarrow M^{\sigma}_{3} \subset L_{2}(\R^3)$, the states in $M^{\sigma}_{3}$ are identified with positions of particles. So the density of states functional $\rho_{t}[\varphi;\psi]$ must be an extension of the usual density of particles $\rho_{t}({\bf a}; {\bf b})$ with initial position ${\bf b}$ in $\R^{3}$. In other words, we must have $\rho_{t}({\bf a}; {\bf b})=\rho_{t}[{\tilde \delta}^{3}_{\bf a}; {\tilde \delta}^{3}_{\bf b}]$.

In the case of macroscopic particles, the conservation of the number of particles is expressed in differential form by the continuity equation. For instance, if $\rho_{t}({\bf a};{\bf b})$ is the density at a point ${\bf a} \in \R^3$ of an ensemble of Brownian particles with initial position near ${\bf b}$ and ${\bf j}_{t}({\bf a}; {\bf b})$ is the current density of the particles at ${\bf a}$, then
\begin{equation}
\label{conti}
\frac{\partial \rho_{t}({\bf a};{\bf b})}{\partial t}+\nabla {\bf j}_{t}({\bf a};{\bf b})=0.
\end{equation}
We will assume that $\rho_{t}({\bf a};{\bf b})$ and ${\bf j}_{t}({\bf a};{\bf b})$ are normalized per one particle, i.e., the densities are divided by the number of particles. In this case, the particle density and the probability density can be identified.

The conservation of states of an ensemble of microscopic particles is expressed by the continuity equation that follows from the Schr{\"o}dinger dynamics. This is the same equation (\ref{conti}) with 
\begin{equation}
\label{contiS}
\rho_{t}=|\psi |^2, \  \ {\rm and } \  \ {\bf j}_{t}=\frac{i\hbar}{2m}(\psi \nabla {\overline \psi}-{\overline \psi}\nabla \psi).
\end{equation}
For the states $\psi \in M^{\sigma}_{3,3}$ we obtain
\begin{equation}
\label{psiB}
{\bf j}_{t}=\frac{\bf p}{m} |\psi|^2={\bf v} \rho_{t}.
\end{equation}
Because the restriction of Schr{\"o}dinger evolution to $M^{\sigma}_{3,3}$ is the corresponding Newtonian evolution,  the function $\rho_{t}$ in (\ref{psiB}) must be the density of particles, denoted earlier by $\rho_{t}({\bf a}; {\bf b})$. Once again, it gives the number of particles that start on a neighborhood of ${\bf b}$ and by the time $t$ reach a neighborhood of ${\bf a}$.
The relation $\rho_{t}({\bf a}; {\bf b})=\rho_{t}[{\tilde \delta}^{3}_{\bf a}; {\tilde \delta}^{3}_{\bf b}]$ tells us that $\rho_{t}$ in (\ref{contiS}) must be then the density of states
$\rho_{t}[{\tilde \delta}^{3}_{\bf a}; \psi]$. It gives the number of particles initially in a state near $\psi$ found under the measurement at time $t$ in the state near ${\tilde \delta}^{3}_{\bf a}$.

We conclude that the flow of states on the space of states contains the flow of particles and the probability flow on $\R^3$ as particular cases. 
However, there is much more to it than just an abstract extension of these physical notions.
For one reason, we saw in the previous section how under a certain random potential associated with a position measurement, the initial state $\psi$ was equally likely to be displaced in any direction on the appropriate projective space of states. As a result, the state was undergoing a random motion on the space of states and the probability to find the state at a point $\varphi$ was shown to be given by the Born rule. 
In terms of the density of states functional, this result can be described as follows: we are dealing with an ensemble of states initially positioned near the point $\psi$ so that the density of states functional is concentrated at the point $\psi$. As the time goes by, the states undergo a random motion in accord with the Schr{\"o}dinger equation with a random potential and the density of states functional ``spreads out" in the space of states. As we saw, the density of states at a point $\varphi$ depends only on the distance from $\psi$ to $\varphi$ and satisfies the Born rule.

Also, from the Schr{\"o}dinger equation and the fact that the Schr{\"o}dinger dynamics constrained to $M^{\sigma}_{3,3}$ is equivalent to the Newtonian one, and using nothing else, we obtained the relationship 
\begin{equation}
\label{density}
\rho_{t}[{\tilde \delta}^{3}_{\bf a}; \psi]=|\psi_{t}({\bf a})|^2. 
\end{equation}
This relationship explains the identification of  $|\psi_{t}({\bf a})|^2$ with the probability density, which is one of the postulates in quantum theory. Indeed, the probability density to find the system in a state for an ensemble of states is proportional to the value of the density of states functional on that state, which for the states in $M^{\sigma}_{3}$ is given by (\ref{density}).
So $|\psi_{t}({\bf a})|^2$ is the probability density to find the particle near ${\bf a}$ simply because this quantity is the density of quantum states near the point ${\tilde \delta}^{3}_{\bf a}$. If there are more states near ${\tilde \delta}^{3}_{\bf a}$, it becomes more likely to find the state under an observation near that point. 

Furthermore, the continuity equation (\ref{conti}) in quantum mechanics follows from the Schr{\"o}dinger equation and is true for {\it any} potential. Suppose we begin with an arbitrary random potential $V$ that under the Newtonian dynamics yields the normal distribution of the position random variable. By section \ref{unique}, there is a unique extension of the Newtonian to Schr{\"o}dinger dynamics. The formula (\ref{density}) asserts then the validity of the Born rule for the resulting distribution of states undergoing the Schr{\"o}dinger evolution with an arbitrary such potential $V$.  
This conclusion extends the results of section \ref{measurements}, originally obtained for the potential typically experienced by the particles in a Brownian motion. 
In addition, a purely geometric derivation of the Born rule in section \ref{Born} acquires here its dynamical validation. Note also that the isotropy of the probability distribution that needed to be assumed in the derivation of section \ref{Born} now follows directly from the Schr{\"o}dinger dynamics and its reduction to the Newtonian one.

It is important to distinguish the deterministic and the stochastic Schr{\"o}dinger evolutions.
The motion of state in quantum mechanics normally follows the deterministic Schr{\"o}dinger equation with a given potential. However, as advocated here, under the conditions typically associated with a measurement, the state evolves by the Schr{\"o}dinger equation with a random potential. The potential initiates a random motion of the state on the space of states and the resulting change in the density functional. The difference between these two types of evolution is analogous to the difference between the usual Newtonian motion of a macroscopic particle in a given potential and the Brownian motion of the particle under random hits, particularly in modeling a measurement by the diffusion. Of course, in light of the discussed relationship of Newtonian and Schr{\"o}dinger dynamics, the analogy is not surprising. 
Note that the typical process of measurement must be sufficiently fast or must satisfy alternative conditions to be able to neglect the deterministic Schr{\"o}dinger evolution during the measurement. These conditions were discussed in section \ref{measurements}. In the opposite case, the motion of the state will consist of the deterministic drift and a random motion about the moving mean. The analogy with the measurement on a macroscopic particle can serve here a guiding principle.

In the integral form, the conservation of states in $L_{2}(\R^3)$ can be written in the following form:
\begin{equation}
\label{Fdiffusion}
\rho_{t+\tau}[\varphi;\psi]=\int \rho_{t}[\varphi+\eta;\psi]\gamma[\eta] D\eta,
\end{equation}
where $\gamma[\eta]$ is the probability functional of the variation $\eta$ in the state $\varphi$ and integration goes over all variations $\eta$ such that $\varphi+\eta \in S^{L_{2}}$. 
When the state of the particle is constrained to $M^{\sigma}_{3}=\R^3$, this equation must imply the usual diffusion on $\R^3$.  
The restriction of (\ref{Fdiffusion}) to $M^{\sigma}_{3}$ means that $\varphi=\tilde{\delta}^{3}_{\bf a}$ and $\eta=\tilde{\delta}^{3}_{{\bf a}+{\bf \epsilon}}-\tilde{\delta}^{3}_{\bf a}$, where ${\bf \epsilon}$ is a displacement vector in $\R^3$. As we already know, the function $\rho_{t}[\tilde{\delta}^{3}_{\bf a};\tilde{\delta}^{3}_{\bf b}]=\rho_{t}({\bf a};{\bf b})$ is the usual density of particles in space.
Let us substitute this into (\ref{Fdiffusion}), replace $\gamma[\eta]$ with the equivalent probability density function 
$\gamma({\bf \epsilon}) \equiv \gamma[\tilde{\delta}^{3}_{{\bf a}+{\bf \epsilon}}-\tilde{\delta}^{3}_{\bf a}]$ and integrate over the space $\R^3$ of all possible vectors ${\bf \epsilon}$. As in the Einstein derivation of the Brownian motion, assume that $\gamma({\bf \epsilon})$  is the same for all ${\bf a}$ and independent of the direction of ${\bf \epsilon}$ (space symmetry). Therefore, the terms $\int \epsilon^{k}\gamma({\bf \epsilon})d{\bf \epsilon}$ and  $\int \epsilon^{k}\epsilon^{l}\gamma({\bf \epsilon})d{\bf \epsilon}$ with $k \neq l$ vanish. It follows as in the Einstein derivation that
\begin{equation}
\label{diffusionR}
\frac{\partial \rho_{t}({\bf a};{\bf b})}{\partial t}=K\Delta \rho_{t}({\bf a};{\bf b}),
\end{equation}
where $K=\frac{1}{2\tau} \int \epsilon^2 \gamma({\bf \epsilon})d{\bf \epsilon}$ is a constant.

The diffusion equation (\ref{diffusionR}) describes the dynamics of an ensemble of particles in the classical space $M^{\sigma}_{3}$. If initially all particles in the ensemble are at the origin, then the density of the particles at a point ${\bf a} \in \R^3$ at time $t$ is given by 
\begin{equation}
\label{nor}
\rho_{t}({\bf a};0)=\left(\frac{1}{4\pi kt}\right)^{\frac{3}{2}}e^{-\frac{{\bf a}^2}{4Kt}}.
\end{equation}
In particular, for the mean-squared displacement of the Brownian particle we obtain the usual
\begin{equation}
\label{x^2}
\frac{d \overline{{\bf a}^2}}{dt}=6K.
\end{equation}

Because the embedding of $M^{\sigma}_3$ into $CP^{L_{2}}$ is isometric, we have ${\bf a}^2=\theta^2$ for small values of the distance $\left\|{\bf a}\right\|_{\R^3}$ (this can be also seen from (\ref{mainO})). Also, the density of particles is equal to the density of states functional constrained to $M^{\sigma}_{3}$. From this and the isotropy of the density of states functional $\rho_{t}[{\tilde \delta}^{3}_{\bf a}; \psi]$ it follows that for the mean-squared displacement of state we must have $\frac{d \overline{\theta^2}}{dt}=const$ for sufficiently small values of $t$. This observation is used in \cite{Kryukov2020} to advance a possible explanation of why the macroscopic particles are constrained to the classical space $M^{\sigma}_{3}$.


\section{Summary and experimental verification}

The dynamics of a classical $n$-particle mechanical system on the classical space $\R^3$ was identified with the Schr{\"o}dinger dynamics with the states constrained to the classical phase space submanifold $M^{\sigma}_{3n,3n}$ in the space of states. Conversely, we saw that there is a unique unitary time evolution on the space of states of a quantum system that yields Newtonian dynamics when constrained to the classical phase space.
This resulted in a tight, previously unnoticed relationship between classical and quantum physics. Under this relationship, the classical Euclidean space $\R^3$ is isometrically embedded into the space of states $CP^{L_{2}}$ with the Fubini-Study metric and is identified with the submanifold $M^{\sigma}_{3}$ of $CP^{L_{2}}$. 
The Newtonian dynamics reigns on $M^{\sigma}_{3}$, while the Schr{\"o}dinger dynamics is its unique extension to the space of states $CP^{L_{2}}$.  Vector fields on $M^{\sigma}_{3}$ have a unique extension to linear vector fields on the space of states. Quantum observables are identified with the associated linear vector fields. Commutators of observables are Lie brackets of the vector fields and are related to the curvature of the space of states. The physical quantities of velocity, acceleration and mass in Newtonian dynamics are now components of the velocity of quantum state.

The process of measurement in quantum mechanics is now an extension of the measurement in classical physics that itself produces a normal distribution of the measured observable and can be described by a diffusion equation. The normal probability distribution on  $M^{\sigma}_{3}$ has a unique extension to $CP^{L_{2}}$ and becomes the   Born rule for the probability of transition between states. The state under a measurement is equally likely to fluctuate in any direction on the space of states. This fact is responsible for the validity of the Born rule for the probability of transition of the initial state. The state is not a cloud in the classical space that somehow ``shrinks" under a measurement. Rather, the state is a point in the space of states that undergoes a random motion with a chance of reaching certain areas of the space in the process. The evolution remains unitary and satisfies the Schr{\"o}dinger equation with a random potential without contradicting the known ``no-go" results (see \cite{Kryukov2020}).

The obtained realization of the Newtonian mechanics in functional terms and the derived relationship of the classical and quantum theories is not just a reformulation of the theory. The results of the classical and quantum mechanics are indeed reproduced in the theory. However, the embedding resulted in a tighter relationship between the theories. This relationship can be experimentally tested. 
A meaningful relationship between Newtonian and Schr{\"o}dinger dynamics can be seen in several places. First, there is a formula (\ref{evolll})  that yields the known result that the speed of evolution of state is equal to the uncertainty in energy, derived in a clear geometrical way. Further, the decomposition (\ref{decomposition}) relates Newtonian velocity and acceleration, and, for the appropriate value of $\sigma$, also the mass of a particle to the corresponding components of the velocity of quantum state. However, these results are consistent with the Schr{\"o}dinger dynamics itself and the Ehrenfest theorem that follows from it and cannot serve a validation of the constructed embedding. 

What helps to come up with an experiment is the ``rigidity" of the embedding: the extension of the Newtonian dynamics and Newtonian models to the space of states is {\it unique}. This allows us to approach the process of measurement in quantum theory in a new way, as an extension of the random motion associated with a classical measurement. An important consequence of this is the notion of a density of state functional and its derived isotropy property that can be tested. Indeed, if several observables are measured on a particular state of a system at the same time, we should be able to test the isotropy of the distribution of frequencies of the measured eigenvalues. That is, the state should be seen transitioning equally frequently to the eigenstates of different observables, positioned at the same Fubini-Study distance from the initial state. The observation of different components of spin of a particle would probably be the easiest way to set up such an experiment. Other possible experimental tests of the proposed embedding of the classical into the quantum are discussed in \cite{Kryukov2020}.

\ack

I would like to thank Thomas Elze, John Klauder and other participants of the DICE2018 workshop, where parts of this paper were presented, for their interest and insightful questions. I am deeply grateful to Larry Horwitz for his faithful support over many years and for the opportunity to speak on the subject at Bar-Ilan and Ariel universities. I would like to thank Thomas Elze for inviting me to give a talk at the DICE2018 and DICE2020 workshop and Martin Land for the invitation to speak at the IARD2018 and IARD2020 conferences. I also want to express my  gratitude to David Schudson for valuable edits and improvements to this paper.


\end{document}